\documentclass[11pt]{article}
\usepackage[dvips]{graphicx}
\usepackage{epsfig}

\newcommand{\be}{\begin{equation}}
\newcommand{\ee}{\end{equation}}
\newcommand{\bel}[1]{\begin{equation}\label{#1}}
\newcommand{\bea}{\begin{eqnarray}}
\newcommand{\eea}{\end{eqnarray}}
\newcommand{\beal}[1]{\begin{eqnarray}\label{#1}}
\newcommand{\nn}{\nonumber}
\newcommand{\nin}{\noindent}
\def\d{\partial}

\def\G{\mathcal{G}}
\def\L{\mathcal{L}}
\def\t{\tilde{t}}
\def\tn{\tilde{n}}
\def\4t{\theta\theta\bar{\theta}\bar{\theta}}
\def\2t{\theta\theta}

\begin{document}
\title{\bf Spontaneous Breaking of Lorentz-Invariance \\
and Gravitons as Goldstone Particles  }

\author{\large{\bf Z.~ Berezhiani $^a$\thanks{\it zurab.berezhiani@aquila.infn.it}
~ and   O.~V. Kancheli $^b$\thanks{\it kancheli@itep.ru}}\\ \\
   $^a${\it Dipartimento di Fisica, Universit\`a di L'Aquila, I-67010 Coppito, AQ, and} \\
   {\it INFN, Laboratori Nazionali del Gran Sasso, I-67010 Assergi, AQ, Italy} \\ \\
   $^b${\it  Institute of Theoretical and Experimental Physics,} \\
   {\it  B. Cheremushkinskaya 25, 117259 Moscow, Russia }
}
\date{}
\maketitle

\begin{abstract}
We consider some aspects of spontaneous breaking of Lorentz
Invariance in field theories, discussing the possibility that the
certain tensor operators may condensate in the ground state in
which case the tensor Goldstone particles would appear.
 We analyze their dynamics and discuss to which extent such a
theory could imitate the gravity. We are also interested if the
universality of coupling of such `gravitons' with other particles
can be achieved in the infrared limit.
 Then we address the more complicated models when such tensor
Goldstones coexist with the usual geometrical gravitons. At the
end we examine the properties of possible cosmological scenarios
in the case of goldstone gravity coexisting with geometrical
gravity.
\end{abstract}
\setcounter{footnote}{0}

\section{ Introduction}

 The idea that the spontaneous breaking of Lorentz Invariance (LI)
is accompanied by the non-scalar Goldstone particles was first
discussed by Bjorken in the seminal paper \cite{Bjorken1}, where
these goldstones where associated with the photons.
 This idea was further extended to nonabelian case and also
\cite{Phillips}-\cite{KrTomb} to tensor condensates and
corresponding goldstones were interpreted as graviton-like
objects.

The possibility to represent some or all gauge fields and
gravitons as the Goldstone particles connected with fluctuations
of vector and tensor condensates can seem quite promising.
Especially this concerns to gravitons, as far as the vector gauge
fields interaction are anyway renormalizable while the bad
behavior of gravity at high virtualities obliges to go to some
different descriptions at small distances - just as in Loop
Quantum Gravity or in String Theory approach.

But it is rather unclear how far can one advance in the direction
of full (or partial) replacing  of geometrical gravity by the composite
goldstone tensor field.
Firstly one must have answers to a number of ``crucial'' questions,
in particular:

\begin{itemize}
\item
is it possible to make the interaction of such a goldstones
universal, as is for a ``usual'' geometrical gravitons of the
general relativity (GR) - this is necessary at least in the
long-distance limit, where the universality of gravitation is
experimentally tested with very high precision;
\item
is it possible to guarantee the Lorentz invariance of
various measurable quantities at  experimentally acceptable level
in the presence of vector and tensor condensates.
\end{itemize}

These questions are not independent. If Lorentz invariance is
restored at large distances, then probably the universality of
interactions will be also restored, because many arguments are
collected \cite{Deser}~(but see also \cite{padman})~ that the only
consistent Lorentz-invariant theory of interacting massless spin
two particles is the general relativity.

Perhaps also the opposite statement is true - if at large
distances the gravitational interactions become universal, then
Lorentz non-invariant terms can die out or turn into gauge fixing
terms, giving no contributions to the measurable quantities.

Unfortunately, there is no understanding if all this is really
true, and there is no more or less reliable model calculations
supporting these hypothesis.

But if these hypotheses are correct  and one the non-universal and
LI-violating contributions to the amplitudes are indeed suppressed
to the ``needed'' level, then such an approach to the gravity
could open up a number of interesting possibilities.
 In particular, one can try to incorporate (unify) the goldstone
gravitons in the standard model  picture, or in some of its
generalization. It seems to us, that remain two main classes of
possible models.

\begin{itemize}
\item
 In the models of the first type there is no geometrical gravity
and the fundamental physics corresponds to some renormalizable
theory in a flat four-dimensional space-time.
 This can be, for example, some supersymmetric grand unified
generalization of the standard model like $SO(10)$, $E(6)$ or so,
supplemented by additional gauge sector that becomes strongly
coupled at some scale $\Lambda$ (related to the Planck scale
$m_{p}\simeq 10^{19}$ GeV) at which the tensor condensates are
formed and thus the Goldstone gravitons emerge.
 In this case no additional dimensions can be introduced without
destroying the renormalizability. An important advantage (or
disadvantage!?) is that for the goldstone-gravity the cosmological
constant is automatically zero \cite{KrTomb}, but perhaps it can
be imitated by the non-LI corrections.

\item
 In second type of models the goldstone-gravity emerging from the
spontaneous breaking of LI coexist with the fundamental
"geometrical" gravity whose scale  $M_P \gg \Lambda$.
 The latter could be originated e.g. by the ``normal'' string
degrees of freedom at very small distances, related to the string
scale $\Lambda_s \sim M_P$, while at much larger distances, or at
energy scale $\Lambda \ll \Lambda_s$, where the field
approximation already works, the condensation of the tensor
operators takes place.
 In this bigravity case we can have two  gravitons: primary
graviton - coming (for example) from the massless modes of
strings, and composite graviton, related to the Goldstone
fluctuations of the tensor condensate.
\end{itemize}

In this paper we consider in details some of these questions and
also discuss the main peculiarities of corresponding cosmological
models. The content of the article is distributed in sections as
follows:

\nin {\bf Section 2}: general structure of tensor condensates in
renormalizable field theories with strong coupling.

\nin {\bf Section 3}:  the Goldstone modes of such condensates and
their effective action.

\nin {\bf Section 4}:  tensor condensates and their fluctuations
in the curved background.

\nin {\bf Section 5}: couplings of the goldstone gravitons to
other particles and  possibility of their universality in the
infrared limit.

\nin {\bf Section 6}: the peculiarities of the tensor condensation
in supersymmetric theories, where the Goldstone supergravity can
appear.

\nin {\bf Section 7}:  cosmological implications of the models
with purely goldstone gravity, and the bigravity models when both
- goldstone and the geometrical gravity are present. In the last
case the geometrical gravity (string) scale can be shifted from
$10^{19}$ GeV to much higher energies  $10^{60}$ GeV.

\nin {\bf Section 8}:  concluding remarks.

\section{\bf Tensor condensates }

In the field theory the (composite) condensates can develop from
the enough strong attraction between virtual particles, so that a
tachyonic bound states with $m^2 < 0$ can be created. As a result
the ground state is coherently filled with such tachyons which in
addition can have non-trivial quantum numbers. In a more
phenomenological approach one can simply forget about this
dynamical stage and directly start from an effective Lagrangian
applicable below some energy scale, where the effective composite
fields get negative $m^2$. The last is a way usually followed when
considering various applications of spontaneous symmetry breaking~
\footnote{In fact, the tachyon formation and condensation does not
necessarily need the strong interaction of virtual particles and
in some cases is possible also in a weak coupling regime. But
the latter usually needs additional symmetry arguments or tuning
and is rather a dynamical exclusion than a general rule. One
possibility is when particles which bound in tachyons are almost
massless, and forces between them are long range. The other known
example is superconductivity, where only the electrons close to
the Fermi surface are essential in pair condensation.}.

The non-abelian gauge theories with adequately restricted
multiplet content usually become strongly coupled below certain
`confinement' scale $\Lambda$, and various composite condensates
can appear, like the quark $\langle \bar{q}q \rangle $ and gluonic
$\langle BB\rangle$ condensates in the QCD.
 Unfortunately, about the formation of condensates  we know mainly
from the experiment (and lattice calculations) and not directly
from theory. Only some special cases, e.g. in gauge theories in
the large $N_c$ limit or in theories with large supersymmetry this
phenomenon can be analyzed in more details.
 We also know (only from the experiment) that in the QCD case the
condensates are Lorentz scalars, and yet we do not understand the
dynamical reasons for this. But may be for other gauge groups with
specific  multiplet contents, different from QCD, the nonscalar
condensates can also appear along with ``usual'' scalar
condensates.

\subsection{\bf Model and general definitions}

Let us consider a theory in a minskowskian space-time based on
some non-abelian gauge symmetry $G_c$ and discuss a general
composite symmetric tensor operator with the ``vacuum'' quantum
numbers~
\footnote{As far as we are in a Minkowski space, the
indices could be moved up and down by means of $\eta_{\mu\nu} =
\eta^{\mu\nu} = {\rm diag}(1,1,1,-1)$, e.g. $  \hat{\tau}_\mu^\nu
= \eta^{\nu\rho} \hat{\tau}_{\rho\mu} $.
}~:
 \bel{oper}
\hat{\tau}_{\mu\nu}  ~=~ {\rm Tr} \big[ \,
         A_1 \cdot \big(\bar{\varphi}\d_{\mu}\d_{\nu} \varphi \big) ~+~
         A_2 \cdot \big( \bar{\chi}(\d_{\mu}\gamma_{\nu} +
         \d_{\nu}\gamma_{\mu}) \chi \big) ~+~
         A_3  \cdot \big(B_{\mu\rho} B_{\lambda\nu} \eta^{\rho\lambda}
         \big) ~+~  \dots  \big]~,
 \ee
where $\varphi$,  $\chi$ and $B_{\mu\nu}$  respectively are the
scalar, fermion and gauge fields of the theory which in the
following will be generically denoted as $\psi$, and the
dimensionfull coefficients  $A_i(\psi)$ are some scalar functions
of the latter fields; for our convenience, these coefficients are
normalized so that the operator $\hat{\tau}_{\mu\nu}$ is
dimensionless, namely $A_i \sim \Lambda^{-4}$ for the dimension 4
terms shown in (\ref{oper}), where $\Lambda$ is a confinement
scale; the trace in (\ref{oper}) is taken over the indices of the
internal gauge group so that all terms are gauge singlets
\footnote{
 For simplicity we do not discuss here a more general case when
$\hat{\tau}_{\mu\nu}$ could have also some nontrivial internal
quantum numbers, so that the condensation of $\hat{\tau}_{\mu\nu}$
would break additionally internal symmetries.}
~of $G$.
 The $\hat{\tau}_{\mu\nu}$ has the same symmetries  as
energy-momentum tensor $T_{\mu\nu}$, and so $\hat{\tau}_{\mu\nu}$
and $T_{\mu\nu}$ can partially enter in the same universality
class with respect to a scale change - therefore on can happen
that in the infrared limit their main parts will become close.

Let us suppose that in certain conditions
the vacuum expectation value of this tensor operator
\be
\langle  \tau_{\mu\nu}(x) \rangle   ~=~
 \int D\psi ~\hat{\tau}_{\mu\nu}~  e^{i\int d^4x L(\psi) }
 \ee
can be nonzero and spontaneously break LI, and let us study the
behavior of the system in a neighborhood of $\langle \tau_{\mu\nu}
\rangle $. To calculate explicitly the effective lagrangian
$\L(t_{\mu\nu}(x))$, representing the response of the system to
the excitation of corresponding degrees of freedom,  especially in
a strong field region, is a rather complicated task, but in
principle there are no principal obstacles.

We illustrate this in a brief formal way.
Let us introduce, as usual, the external current $J_{\mu\nu}(x)$
linearly coupled with $\tau_{\mu\nu}(x)$.
Then the response of the system to the excitation of the degrees
of freedom related to $\tau_{\mu\nu}(x)$ can be described by the
generating functional
 \beal{genfun1}
& Z[J_{\mu\nu}(x)] ~=~ \int D\psi \,
e^{i\int d^4x [L(\psi) +\tau_{\mu\nu}(\Phi) J^{\mu\nu}(x) ] } =
\nn \\
&  =~ \int D\psi \, Dt_{\mu\nu} \, \delta_x [t_{\mu\nu}(x)-
 \tau_{\mu\nu}(\psi(x)) ] \,
~e^{i\int d^4x [L(\psi) ~+~ t_{\mu\nu}(x)J^{\mu\nu}(x)] }~=
\nn\\
& ~=~ \int  D t_{\mu\nu} \, e^{ i\int d^4x [\L (t_{\mu\nu}) ~+~
        t_{\mu\nu} J^{\mu\nu} ]  }~~,~~~~~~~~~~~~~~~
\eea
where we introduced the integration with the $\delta_x$ functional
over the auxiliary field $t_{\mu\nu}(x)$ connected with the operator
$\tau_{\mu\nu}$.
In this way, by integrating out the fundamental degrees of freedom with
the constraint,  one can define
the effective lagrangian for the effective tensor field
$t_{\mu\nu}$:
\bel{genfun2}
i \int d^4x \L(t_{\mu\nu}(x)) ~=~  \ln \Big[ \int D\psi \,
e^{i\int d^4x L(\psi)}
 ~\delta_x [t_{\mu\nu}(x)- \tau_{\mu\nu}(\psi)] \Big] ,
\ee
As far as we start with initially Lorentz-invariant theory, the effective
Lagrangian $\L(t_{\mu\nu})$ must be Lorentz-invariant.
Next, we suppose that for a slowly changing $t_{\mu\nu}(x)$
it is approximately local and can be expanded
in powers of $t_{\mu\nu}$ and its derivatives
$t_{\mu\nu ,\alpha}(x)=\d_\alpha t_{\mu\nu}$:
 \bel{lagran}
\L(t_{\mu\nu}) ~=~ - V(t_{\mu\nu}) ~+~
  \Gamma^{\alpha\beta\gamma\delta\rho\sigma}
  t_{\alpha\beta , \gamma} t_{\delta\rho , \sigma} ~+~
  W^{\alpha\beta\gamma\delta\rho\sigma\lambda\varsigma\tau}
  t_{\alpha\beta , \gamma} t_{\delta\rho , \sigma}
  t_{\lambda\varsigma, \tau} ~+~ ...
\ee
where generically also the coefficients $\Gamma , W, ...$
are functions of the field $t_{\mu\nu}(x)$, while
in most general case the scalar potential  $V(t_{\mu\nu})$
is a function of four independent Lorentz-invariant combinations:
 \bel{s}
s_1 = {\rm Tr}(t)    ~~~
s_2 = {\rm Tr}(t^2)  , ~~~
s_3 = {\rm Tr}(t^3)  , ~~~
s_4 = {\rm Tr}(t^4)
 \ee

Therefore,  we are interested in a situation when
near the confinement scale $\Lambda_c$
the nontrivial vacuum expectation value can be generated,
\bel{mean}
\langle \hat{\tau}_{\mu\nu} \rangle ~=~  \langle t_{\mu\nu} \rangle  ~=~
 n_{\mu\nu}~,
\ee
where $n_{\mu\nu}$ is a constant symmetric tensor
with the elements of the order of 1.

The general form of potential $V$ for not very large $t_{\mu\nu}$ is
\footnote{Here we only supposed that $V(t_{\mu\nu})$ is analytic at
$t_{\mu\nu} = 0$. We also symbolically denoted as $t_{\mu\nu}^n$
invariants, containing n-th power of $t_{\mu\nu}$, for example
$t_{\mu\nu}^2 =t_{\mu\nu}t_{\alpha\beta}\eta^{\mu\alpha}\eta^{\nu\beta}$,
where $\eta_{\alpha\beta}$ is Minkovski metric.
}
 \bel{poten}
 V(t_\mu^\nu) ~=~ V_0 ~+~ g_1 t_\mu^\mu  ~+~
 g_2 t_\mu^\nu t^\mu_\nu  ~+~ g_3 t_\mu^\nu t_\nu^\rho t_\rho^\mu
 ~+~ g_4 t_ \mu^\nu t_\nu^\rho t_\rho^\sigma t_\sigma^\mu ~+~ \cdots ~,
\ee
where generally the dimensional coefficients $g_i$  are
expected to be related to the confinement scale $\Lambda$ as
$g_i \sim \Lambda^{4}$.
As far as $t_{\mu\nu}^2$ is not positively defined such potentials
can easily lead to the nontrivial minima for very broad area of
coefficients~$g_i$.
Therefore, one can expect that near to the confinement scale
$\Lambda_c$ the nontrivial expectation value can be generated,
\bel{mean}
\langle \hat{\tau}_{\mu\nu} \rangle ~=~  \langle t_{\mu\nu} \rangle  ~=~
n_{\mu\nu}~,
\ee
where $n_{\mu\nu}$ is some constant symmetrical tensor
($n_{\mu\nu} \neq \eta_{\mu\nu}$) with the elements of the order~of~1.

In fact we will take later in mind a slightly more general
effective lagrangian $\L(t_{\mu\nu},\phi_i)$~, containing also the
fields $\phi_i$ of the ``standard model'' (or its generalization),
coupled to $\psi=(B,\chi,\varphi)$ fields by some mediators. We also
suppose that these couplings are rather weak, so that on can
neglect the contributions from $\phi_i$ fields to the
$(B,\chi,\varphi)$ dynamics.

\subsection{\bf Configurations of the tensor condensate}
The possible configurations of $n_{\mu\nu}$
can be classified in terms of four eigenvalues $\lambda^{(a)}$:
\bel{ev}
 n_{\mu\nu} S_{\nu}^{(a)} ~=~ \lambda^{(a)} S_{\mu}^{(a)}~,
\ee
where $ S_{\mu}^{(a)}$ are the orthonormal eigenvectors:
$$
S_{\mu}^{(a)} S_{\mu}^{(b)}~=~\delta^{ab},~~~~~~
\sum_a S_{\mu}^{(a)} S_{\nu}^{(a)}~=~\eta_{\mu\nu},~~~~~~
\sum_a S_{\mu}^{(a)} S_{\nu}^{(a)} \lambda^{(a)}~=~n_{\mu\nu}~~.
$$
This means that by a rotations and busts on can chose such a
coordinate system in with $n_{\mu\nu}$ is diagonal
\bel{n1}
n_{\mu\nu} ~ = ~
\left(%
\begin{array}{cccc}
  \lambda_1 & 0 & 0 & 0 \\
  0 & \lambda_2 & 0 & 0 \\
  0 & 0 & \lambda_3 & 0 \\
  0 & 0 & 0 & -\lambda_0 \\
\end{array}%
\right)~~~,
\ee
or it has one of the possible `light-like' forms
\bel{n2}
n_{\mu\nu} ~ = ~
\left(%
\begin{array}{cccc}
  \lambda_1 & 0 & 0 & 0 \\
  0 & \lambda_2 & 0 & 0 \\
  0 & 0 & \lambda_4 & \lambda_5 \\
  0 & 0 & \lambda_5 &  -\lambda_4 \\
\end{array}%
\right) ~ ,  ~~~~~
n_{\mu\nu} ~ = ~
\left(%
\begin{array}{cccc}
  \lambda_1 & 0 & 0 & 0 \\
  0 & \lambda_2 & 0 & 0 \\
  0 & 0 & \lambda_4-\lambda_5 & \lambda_5 \\
  0 & 0 & \lambda_5& -\lambda_4 -\lambda_5 \\
\end{array}%
\right)
\ee
In the light-like cases we will always have the preferential
space direction.
For the diagonal $n_{\mu\nu}$ the eigenvectors in this frame are
also diagonal $S^a_{\mu}~\sim~\delta^a_{\mu}~\lambda^a$, and for
`light-like' $n_{\mu\nu}$ the eigenvectors $S^a_{\mu}$ can have
also two nondiagonal components - for example $S^3_0$ and $S^0_3$.

In the following, we consider only first case, with the diagonal
$n_{\mu\nu}$ in the form (\ref{n1}) with real eigenvalues
$\lambda^{(a)}$. The eigenvectors in this frame are also diagonal:
$S^a_{\mu}~\sim~\delta^a_{\mu}~\lambda^a$,

Clearly, we are not interested in the case when LI is unbroken, i.e.
$n_{\mu\nu} \propto  \eta_{\mu\nu}$. The LI breaking needs that
the irreducible "traceless" part of $n_{\mu\nu}$,
$\bar{n}_{\mu\nu}$ is non-zero.

As far as we start with initially Lorentz-invariant theory, in
most general case the scalar potential $V(t_{\mu\nu})$ is a
function of four independent Lorentz-invariant combinations:
\bel{orpot}
V(t_{\mu\nu}) ~=~ V(s_1, s_2, s_3, s_4)
\ee
of four independent invariants
\bel{invs}
s_1 = {\rm Tr}(t) = t_\mu^\mu,  ~~~
s_2 = {\rm Tr}(t^2) = t_\mu^\nu t^\mu_\nu , ~~~
s_3 = {\rm Tr}(t^3) = t_\mu^\nu t_\nu^\rho t_\rho^\mu , ~~~
s_4 = {\rm Tr}(t^4) =  t_ \mu^\nu t_\nu^\rho t_\rho^\sigma t_\sigma^\mu
 \ee
Therefore the the conditions for extremum $\d V/\d t_{\mu\nu} = 0$
are reduced to $\d V/\d s_i = 0$~ and the stability conditions for
condensate is  ~$\d^2 V/\d s_i~\d s_k~>~0$.~ The roots
$\tilde{s}_i$ of equation $\d V/\d s_i = 0$ are related to the
eigenvalues of matrix~$n_{\mu\nu}$~:

\bel{roots} \tilde{s}_1 =\sum_{i=1}^4 \lambda_i~,~~~ \tilde{s}_2
=\sum_{i=1}^4 \lambda_i^2~,~~~ \tilde{s}_3 =\sum_{i=1}^4
\lambda_i^3~,~~~ \tilde{s}_4 =\sum_{i=1}^4 \lambda_i^4~~. \ee The
simplest model example of potential, such as
$$
 V(s_i) ~=~ v_0 \sum_{i=1}^4 (-a_i s_i + \frac{1}{2} s_i^2~)~~,
$$
already leads to general form of spontaneous LI breaking.
In this case the roots, coming from conditions ~$\d V/\d s_i = 0$,
are
$$
s_i = a_i ~,~~~ V(a_i) = - \frac{1}{2} v_0 a_i ~,~~~
\frac{\d^2 V(a_i)}{\d s_i^2} = v_0~~,
$$
and, depending from values of $a_i$, on can have configurations
(\ref{n1}) or (\ref{n2}) with all different or equal $\lambda_i$.

The degenerate case, when some of eigenvalues $\lambda_i$ may
coincide, can lead to number of  simplifications. For example the
case of maximal degeneracy $\lambda_1=\lambda_2=\lambda_3 \neq
\lambda_0$. In this case on can go to a system where the space is
isotropic and this will lead to additional restrictions on
possible goldstone-gravity configurations (see next section). But
there is no reason to expect such a degeneracy for the general
potentials  $V(s_i)$, although on can meet the approximate
degeneracy if the Lorentz-invariance is restored at long
distances.

When LI is spontaneously broken and $<t_{\mu\nu}> \neq 0$,
the mean value of energy-momentum tensor  for the $\phi_i$
fields in vacuum can be also nontrivial and has the general form~:
\bel{tmn}
 \langle ~T_{\mu\nu} \rangle ~=~
a_0 \eta_{\mu\nu} ~+~ a_1 n_{\mu\nu} ~+~
  a_2n_{\mu\gamma}n_{\gamma\nu} ~+~ \cdots
\ee
In the Lorentz frames, where the $n_{\mu\nu}$ is diagonal, the
$T_{\mu\nu}$ is also diagonal.
In this case the $n_{\mu\nu}$ condensate can be interpreted
as an anisotropic ``solid media'' filling the space
with the energy density $T_{00}$ and anisotropic pressures $T_{kk}$.

The $a_0$ term in (\ref{tmn}) is usually  perturbatively divergent
in renormalizable theories, and this leads to problems in GR, but
here this  Lorentz-invariant part $\sim  \eta_{\mu\nu}$ can not be
taken into account (it can be freely subtracted)
\footnote{As is usually done in field theories without gravity
because it does not affect any measurable quantity.
},
because it does not interact with the goldstone gravitons. The
non-LI contributions in $T_{\mu\nu}$ are probably nonuniversal and
the coefficients $a_i$ are scale-dependent. Near the confinement
scale  $a_i \sim \Lambda^4$, but they, can become small on large
distances as e.g. in $\G_{\mu\nu}^{(i)}$ (\ref{covar}).

If usual general-relativistic geometrical gravity is also added to
the renormalizable theory with $\psi=(B,\chi,\varphi)$ fields,
then the big mean invariant $T_{\mu\nu}$ can lead to unacceptable
cosmological behavior. But at the same time here we have
additional possibility for solving cosmological ``constant''
problem, because on can essentially restrict the strength of
interaction of the geometrical gravitons with $\langle T_{\mu\nu}
\rangle$ (up to the level of today's value of cosmological
constant) ~(See section 6.).

\subsection{\bf LI violations in observable}
The constant tensor $n_{\mu\nu}$ can enter in various observable
quantities, and we will interpret this as a violation of Lorentz
invariance (LI).
In general, all Green functions and scattering amplitudes
may depend from $n_{\mu\nu}$.
At the scales $|x_i -x_j| \gg \Lambda^{-1}$ the condensate $n_{\mu\nu}$
can be considered as a constant external tensor field.
Then for the external momenta $k_i \ll \Lambda$ the amplitudes
\bel{aml}
 A_m(k_1, k_2, ..., k_m,~ n_{\mu\nu} )
\ee
can be calculated from some effective low-energy Lagrangian of the
``physical" particle fields $\phi$ which can include the external
``spurion'' field $n_{\mu\nu}$  via various vertexes:
\bel{vert}
\phi\d_{\mu}\d_{\nu}\phi ~n_{\mu\nu}~,~~~
\phi^k \d_{\mu}\phi \d_{\nu}\phi ~n_{\mu\nu}~,~~~
\phi\d_{\mu}\d_{\nu}\phi ~n_{\mu\lambda} n_{\lambda\nu}  ~,\cdots
\ee
Then the $\phi$-particles propagators get tree level corrections
depending on $n_{\mu\nu}$ which can be expressed as
\beal{prop}
G^{-1}(k) ~=~ k^2 - m^2 + ~k_{\mu} k_{\nu}N_{\mu\nu} ~+~ O(k^4)
~=~ \mathcal{G}_{\mu\nu} k_{\mu} k_{\nu}- m^2 ~+~ O(k^4)  \\
\mathcal{G}_{\mu\nu} ~=~ \eta_{\mu\nu} + N_{\mu\nu}~,~~~~~~
N_{\mu\nu} = c_1 n_{\mu\nu} + c_2 n_{\mu\lambda }n_{\lambda\nu}
 ~+\cdots~, \nn
\eea
where coefficients $c_i$ are proportional to coupling constants,
with which operators (\ref{vert}) enter the effective lagrangian.
The term of type (\ref{vert}) come also from expansion of the
vertices with loops in external momenta. There are also induced
vertices of same general type (\ref{vert}), where some
$n_{\mu\nu}$  are replaced by Minkowski metric $\eta_{\mu\nu}$.

So these vertexes can also be represented in combined form
\bel{vert2}
\phi\d_{\mu}\d_{\nu}\phi ~\G_{\mu\nu}^{(1)}~,~~~
\phi^k \d_{\mu}\phi \d_{\nu}\phi ~\G_{\mu\nu}^{(2)}~,~~~
\phi\d_{\mu}\d_{\nu}\phi ~\G_{\mu\lambda}^{(3)}
\G_{\lambda\nu}^{(4)}  ~,\cdots
\ee
where
$$
\G_{\mu\nu}^{(i)} ~=~ c^{(i)}_1 \eta_{\mu\nu} +
  c^{(i)}_2 n_{\mu\nu} ~+~\cdots
$$
Now, if coefficients  $\G_{\mu\nu}^{(i)}$ are
expected to be universal,
they all should be approximately the same
$$
\G_{\mu\nu} ~=~ \G_{\mu\nu}^{(1)} ~=~
 \G_{\mu\nu}^{(2)} =~\cdots~~~,
$$
and hence  $\G_{\mu\nu}$ can be interpreted as a new metric
corresponding to flat space with non-orthogonal coordinates~
\footnote{
Even for slightly nonuniversal $\G_{\mu\nu}^{(i)}$ we can have the
nonstandard $(\varepsilon,\vec{k})$ dispersion relations like
$\varepsilon^2 = \gamma_i k^2 +m^2,...$, where $\gamma_i$ depends
from particles type. This can lead to Cherenkov radiation in
vacuum and to other ``kinematically forbidden'' process. All these
topics is widely discussed in last years from various
angles~\cite{colgle,jacob},~ in particular in connection with GKZ
cutoff.
}.

This geometrical interpretation can be simply generalized to
$n_{\mu\nu}~\rightarrow t_{\mu\nu}(x)$
slowly varying with $x$ on the scale $\Lambda^{-1}$.

But in general the coefficients $\G_{\mu\nu}^{(i)}$
are not universal, at least near to the scale $\Lambda$,
and they depend on the particle $\phi$ and vertex type.
So it is probably impossible to interpret the additional terms
in (\ref{prop}) and (\ref{vert})  as coming from motion of
$\phi$ particles in nonorthogonal coordinates, represented by
pure gauge gravitational field  $\G_{\mu\nu}$,
if this motion is \underline{``measured'' at distances $\sim \Lambda^{-1}$.}
For such measurements vacuum will look like a highly anisotropic crystal.

But it is possible that at  large distances $r \gg \Lambda^{-1}$
the difference between various $\G_{\mu\nu}^{(i)}$
can become very small. This can be represented as:
\bel{covar}
\G_{\mu\nu}^{(i)} ~=~ \G_{\mu\nu} ~+~
   \frac{k^2}{\Lambda^2} ~Z_{\mu\nu}^{(i)} ~+~
   \frac{k_{\alpha} k_{\beta}n_{\alpha\beta}}{\Lambda^2}
     ~Y_{\mu\nu}^{(i)} ~+~ ...~~~,
\ee
where the nonuniversal contributions to $\G_{\mu\nu}^{(i)}$
are strongly suppressed close to ``our'' scale
\footnote{
Note that from experiment we have rather strong limitations on such
a mean $|\delta G_{\mu\nu}^{(i)}|_{exper.} <  10^{-15} \div 10^{-20}$,
but which can be covered by the factor  $ k^2/\Lambda^2$~.
}.
We consider this possibility in more details in Section 5.

\section{\bf Tensor condensate oscillations}

\subsection{\bf Goldstone gravitons}
So let us suppose that the $(B,\chi,\varphi)$ field system is such
that the condensate $\langle \tau_{\mu\nu} \rangle $ is generated.
As discussed in previous section this means that the minimum of
potential $V(t_{\mu\nu})$~, defined by the condition $\d V / \d
t_{\mu\nu} = 0$ is in the nontrivial point $\langle t_{\mu\nu}
\rangle  = n_{\mu\nu} = \eta_{\mu\nu}+ \tn_{\mu\nu}$.

Further, because $V(t_{\mu\nu})$  depends only from invariants
formed from  $t_{\mu\nu}$ and by definition does not depend on the
derivatives of  $t_{\mu\nu}$ over $x$, the field configurations
$n_{\mu\nu}(x)$ which  can be obtained from $n_{\mu\nu}$ by the
arbitrary local Lorentz  transformation
\bel{omen}
  t_{\mu\nu}(x) ~=~
 \Omega_{\mu}^{\lambda}(x) ~n_{\lambda\sigma}~ \Omega_{\nu}^{\sigma}(x)
 ~= ~ \eta_{\mu\nu} ~ + ~
 \Omega_{\mu}^{\lambda}(x) ~\tn_{\lambda\sigma}~ \Omega_{\nu}^{\sigma}(x),
\ee
give the same value of the potential as $n_{\mu\nu}$:
~$V(t_{\mu\nu}(x))~=~V(n_{\mu\nu})$.
Thus, the variables $\Omega_{\mu}^{\nu}(x)$ are connected with the
massless Goldstone-like degrees of freedom,
whose Lagrangian are given only by the terms with derivatives
$\d_{\lambda} t_{\mu\nu}(x) $ in (\ref{lagran}),
while they have no interaction potential~
\footnote{
As far as all loop corrections due to fundamental fields $\psi$ are
already included in the effective potential $V(t_{\mu\nu})$,
we do not need to care about corrections
to $V(t_{\mu\nu})$ from the Goldstone loops.}.
The local $O(3,1)$ rotation matrixes
\bel{omdef}
\Omega_{\mu}^{\nu}(x) ~=~ \big( \exp [ \frac12\omega_{a b}(x) \Sigma^{a b} ]
\big)_{\mu}^{\nu}
\ee
depend on antisymmetric ``angular'' fields ~$\omega_{a b}$(x)~
which represent six independent flat directions.
Here $\Sigma^{ab}$ is the spin part of generators of
Lorentz rotation in the vector representation:
$$
\big( \Sigma^{a b} \big)^{\nu}_{\mu} ~=~ \eta^{a \nu} \delta^{b}_{\mu}
-  \eta^{b \nu} \delta^{a}_{\mu}~~,
$$
where the local $(a,b)$ and tensor $(\mu,\nu)$ indices are `mixed'.
The full infinitesimal action of these operators on a tensor,
$ \big( \Sigma_{\varrho\sigma} \big)^{\alpha\beta}_{\mu\nu}
t_{\alpha\beta}$, is given by relation
$$
\big( \Sigma_{\varrho\sigma} \big)^{\alpha\beta}_{\mu\nu} ~=~
 \eta_{\sigma\mu}~\delta^{\alpha}_{\rho}~\delta^{\beta}_{\nu} ~-~
      \eta_{\rho\mu}~\delta^{\alpha}_{\sigma}~\delta^{\beta}_{\nu} ~+~
      \eta_{\sigma\nu}~\delta^{\alpha}_{\mu}~\delta^{\beta}_{\rho} ~-~
      \eta_{\rho\nu}~\delta^{\alpha}_{\mu}~\delta^{\beta}_{\sigma}~.
$$
Therefore, taking all this into account, one can divide dynamical variables
$t_{\mu\nu}(x)$ in the massless Goldstone-like ``rotational" modes
connected with  $\Omega_{\mu}^{\lambda}(x)$ and the massive modes
$\varrho^a(x)$ connected
\footnote{
The masses of the fields $\varrho^a(x)$ are defined by the second
derivatives $\d^2 V/ \d t_{\mu\nu}^2$ and in general are expected
to be order $\Lambda$. Note that  one should not expect that they
correspond directly to some new massive stable particles formed on
the scale $\Lambda$. They  contain the operator combinations of
the fundamental fields $\psi = B,\chi,\varphi$ in approximately
the same proportion as in the decomposition~(\ref{oper}) for
$\tau_{\mu\nu}$, and can be highly unstable.
}
with the excitations of the eigenvalues $\lambda^a$~:
\bel{decomp}
t_{\mu\nu}(x) ~=~ \sum_{a=1}^4 ~w_{\mu}^{a}(x)~
\big( \lambda^{(a)} + \varrho^a(x)\big)~w_{\nu}^{a}(x)~~,
\ee
where
$w_{\mu}^{a}(x) ~=~ \Omega_{\mu}^{\lambda}(x) S_{\lambda}^{a}$
~are the analogues of local tetrad vectors, often used for representing
the gravitational field.
Here these tetrads $w_{\mu}^{a}(x)$ are restricted
by a ``gauge'' condition that in every point of the space-time they
can be reduced by the local Lorentz-transformation to the predefined
constant vectors $S_{\lambda}^{a}$.

One can also represent (\ref{decomp}) in a slightly different form
\bel{tomd}
t_{\mu\nu}(x) ~=~ \Omega_{\mu}^{\lambda}(x)~
\big[\tn_{\mu\nu} + \t_{\lambda\sigma}(x)\big]
   ~\Omega^{\sigma}_{\nu}(x)~~,
\ee
where $\tn_{\mu\nu}$ is a traceless part of  $n_{\mu\nu}$
($\eta^{\mu\nu}\tn_{\mu\nu} = 0$) and
\bel{tns}
\t_{\mu\nu}(x) ~=~
~  S_{\mu}^{a}~S_{\nu}^{a}~\varrho^a(x)
\ee
is a part of $t_{\mu\nu}(x)$ that contains only the massive modes
while the massless modes are encoded in $\Omega_{\mu}^{\lambda}$.
(Here and below the summation over the ``eigenvalue''
index $a$ is assumed as in (\ref{decomp}).)
 For small $\omega_{a b}(x)$ the massless fields
($\omega_{\mu\nu} = -\omega_{\nu\mu}$)
enter in linear form
$\Omega_{\mu\nu}(x) = \eta_{\mu \nu} + \omega_{\mu\nu}$,
and hence the week field decomposition of $t_{\mu\nu}(x)$ in massless
fields is given by the series
\bel{tdec}
t_{\mu\nu}(x) ~\simeq~   n_{\mu\nu}(x) ~+~
\omega_{\mu\gamma}(x)~ \tn_{\gamma\nu}(x)~+~
\tn_{\mu\gamma}(x)~\omega_{\gamma\nu}(x)~+~\cdots
\ee
So the symmetric tensor $h_{\mu\nu}(x)$ entering in
usual linearized description of weak gravitational field
$$
t_{\mu\nu}(x) ~\simeq~ n_{\mu\nu} ~+~ h_{\mu\nu}(x)~,~~~~~~
|h_{\mu\nu}| \ll |n_{\mu\nu}|~,
$$
can be represented by antisymmetric goldstone fields
$\omega_{\alpha\beta}$
\bel{homega}
h_{\mu\nu}~\simeq~\omega_{\mu\beta} \tn^\beta_\nu ~-~
    \omega_{\nu\beta} \tn^\beta_\mu~,~~~~~
    \omega_{ab}~\simeq~h_{a\beta} \tilde{n}_{\beta b} ~-~
    h_{b \beta} \tilde{n}_{\beta a}~,
\ee
where $\tilde{n}_{\alpha\beta}  = (n^{-1})_{\alpha\beta} \simeq
n_{\alpha\beta}$ is function
 of $n_{\alpha\beta}$.
In fact this representation of $h_{\mu\nu}$   in terns
of $\omega_{\mu\nu}$ corresponds to a definite gauge choice in linear
approximation~
\footnote{
Note, that this gauge in which the goldstone gravitons appear
is similar to the axial gauge $n_{\mu} A^{\mu} = 0$ for vector
goldstone particles, where $n_{\mu}$ is proportional to the
value of vector condensate}.
The general ``gauge fixing'' corresponding to graviton-goldstones
is given in (\ref{gaucon}).

\subsection{\bf Lagrangian}
Between the degrees of freedom contained in $t_{\mu\nu}(x)$
in the neighborhood of $n_{\mu\nu}$ should be no
tachyons, as far we are already in the minimum of  $V(t)$ with
respect to all independent variations of components of
$t_{\mu\nu}$. One should not expect any ghosts
contributions in terms with derivatives~ $t_{\alpha\beta ,
\gamma}$, so that all signs of corresponding kinetic terms in
(\ref{lagran}) should be correct -- otherwise there will be a
noncausal propagation of $t_{\mu\nu}$ fields, but it is impossible
in terms of primary fields $(B,\chi,\varphi)$. These conditions
restricts the form of coefficients
$\Gamma^{\alpha\beta\gamma\delta\rho\sigma}(t)$~, ...~,~ entering
effective lagrangian (\ref{lagran}) near the $t_{\mu\nu} =
n_{\mu\nu}$. And just this second term
 \bel{secterm}
\L_2  ~=~ \Gamma^{\alpha\beta\gamma\delta\rho\sigma}
  t_{\alpha\beta , \gamma} t_{\delta\rho , \sigma}~~,
\ee
in the lagrangian (\ref{lagran}) defines the minimal dynamics
for the goldstone fields $\omega_{\mu\nu}$. The stability
conditions for $n_{\mu\nu}$ can be also be formulated in terms of
 corresponding effective Hamiltonian
$$
\mathcal{H} ~=~ t_{\alpha\beta , 0}~ \d \mathcal{L}_2
  /\d(t_{\alpha\beta , 0}) ~-~ \L_2~.
$$
For this one must require that for all small variations
~$|\omega_{\alpha\beta}| \ll1$~, ~$|\varrho^a| \ll\Lambda$
of  $t_{\mu\nu}$ fields
\be
\delta t_{\mu\nu}(x) ~\simeq~  \varrho^a(x)~ S^{a}_{\alpha}S^{a}_{\nu}
~+~ \omega_{\alpha\beta}(x)~ \big(
g_{\mu\alpha} n_{\nu\beta} ~-~ n_{\mu\alpha} g_{\nu\beta} ~+~
g_{\nu\alpha} n_{\mu\beta} ~-~ n_{\nu\alpha} g_{\mu\beta}
\big)
\ee
the Hamiltonian remains positively defined, and the corresponding
variation $\delta \mathcal{H} \ge 0$ in the neighborhood of
$t_{\mu\nu}(x) = n_{\mu\nu}$.

The equation of motion for the Goldstone field
$\delta \mathcal{L}/ \delta\omega_{ab}(x)$ = 0~ follow from
the terms with derivatives in (\ref{lagran}).
On the other hand these equations must be also contained in the
conservation lows
$\d_{\mu} J_{[\alpha\beta]\mu} =0$ for the Noetter currents,
connected with the symmetry which is spontaneously broken.
In our case these currents are the full angular momentum densities
$J_{[\alpha\beta]\mu} $ for field system  $\psi=(B,\chi,\varphi)$,
or for  lagrangian (\ref{lagran})  for the effective fields
$t_{\mu\nu}$.
Since the Lagrangians are translation invariant these
equations reduce to
$$
 T_{\alpha\beta}  ~-~ T_{\beta\alpha}   ~-~ \d_{\mu} S_{[\alpha\beta]\mu}
 ~= ~ 0 ~,
$$
where  $T_{\alpha\beta}$ is canonical  energy-momentum tensor,
and $S_{\alpha\beta\mu}$ - is the spin part of  $J_{[\alpha\beta]\mu}$.

To find the form  of the Lagrangian $\mathcal{L}_2$
for weak fields, including also their minimal interactions, we need
the derivatives of $t_{\alpha\beta}$ up to the second order in
physical fields
\beal{tdec1}
\d_{\lambda} t_{\mu\nu} ~\simeq~
\d_{\lambda}\varrho^a \Big( S^{a}_{\mu} S^{a}_{\nu} ~+~
\omega_{\mu\alpha} S^{a}_{\alpha}S^{a}_{\nu} ~+~
\omega_{\nu\alpha} S^{a}_{\alpha}S^{a}_{\mu} \Big) ~+~  \\
~+~ \d_{\lambda} \omega_{\alpha\beta}~ \big(
g_{\mu\alpha} \t_{\nu\beta} ~+~
g_{\nu\alpha} \t_{\mu\beta} ~+~
n_{\mu\alpha} \omega_{\beta\nu}  + n_{\nu\alpha} \omega_{\beta\mu}
\big)~+ \cdots~~~~~~   \nn
\eea
In the week field approximation tensor $\Gamma^{......}$
can be taken as a constant, depending only from
$\eta_{ik}$ and $n_{ik}$.
Then the general form of the Lagrangian $\mathcal{L}_2$
(\ref{secterm}) reads as
\beal{aaa}
\mathcal{L}_2  ~\simeq~
\big( P^{\gamma\alpha\beta\sigma\delta\rho} +
\omega_{ab} Q^{ab~\gamma\alpha\beta\sigma\delta\rho} \big)
\d_{\gamma} \t_{\alpha\beta}
\d_{\sigma} \t_{\delta\rho} ~+~~~~~~~~~~  \nn    \\
\big( \d_{\gamma} \omega_{ab}
\d_{\sigma} \omega_{mn}\big)
\big( \t_{\alpha\beta} \t_{\delta\rho} \big)
\big(  H_1^{\gamma a b~ \cdots~  \rho}  + \omega_{..}H_2^{.......} \big)  ~+~~~            \\
\d_{\gamma} \omega_{ab}\cdot \big(
\t_{\alpha\beta}   \d_{\gamma} \t_{\delta\rho}\big)
U^{\gamma ab~\alpha ~\cdots~ \rho} ~+ \cdots~, \nn
\eea
where $P^{\cdots} , Q^{\cdots}, H^{\cdots}$ are constant tensors,
constructed from $\eta^{ik}$ and $n^{ik}$, so that to fulfill
the $\mathcal{H}$ stability conditions.
In (\ref{aaa}) the first and second lines correspond to kinetic
terms for the massive and goldstone modes and the third line
describes their interaction in which, as is usual for goldstone
particles,  fields $\omega_{\alpha\beta}$ enter only through the
derivative terms~
\footnote{ Notice that the states described by $\omega_{ab}$ and
$\t_{\alpha\beta}$ are orthogonal and so there should be no direct
mixing between them, i.e. no terms of the type ~$\d_\mu
\omega_{ab} ~\d_\mu t_{ab} $.
}.

 It is very essential and specific for the non-scalar goldstone
particles that the massless fields $\omega_{\alpha\beta}$ can
enter also in the coefficient of the kinetic term for massive mode
without derivatives (first line in (\ref{aaa})).
 For the scalar goldstones such terms are cancelled in all orders
and the goldstone interaction with other fields enter only through
the derivative terms - like in the third line in  (\ref{aaa}))~
\footnote{ For rather general $L=\d_{\mu}\phi^{\dag}
\d_{\mu}\phi-V(\phi)$ where $\phi$ - arbitrary column vector we
can localize $\Omega$ - the global symmetry of $V(\phi)$
separating the massive $\rho$ and the ``massless'' modes
$\d_{\mu}(\Omega\rho)^{\dag}\d_{\mu}(\Omega\rho)$ where $\Omega(x)
=\exp(i\vec{T} \vec{\omega}(x))$.
 For $<\rho> \neq 0$ the goldstone fields $\vec{\omega}(x)$ fully
decouple from massive kinetic term and  enter in L only trough
derivative $L = \d_{\mu}\rho^{\dag}\d_{\mu}\rho + (\rho^{\dag}\rho
) \d_{\mu}\vec{\omega}\d_{\mu}\vec{\omega} +
\d_{\mu}\vec{\omega}\vec{J}_{\mu} - V(\rho)$ where  the current
$\vec{J}_{\mu} = i(\d_{\mu}\rho^{\dag}\vec{T}\rho -
\rho^{\dag}\vec{T}\d_{\mu}\rho)$.
 So the interaction of such a $\vec{\omega}$-goldstones is switched
off when their momenta is going to zero. The `diagonal' terms in
$\mathcal{L}_2$ of type $\d_{\lambda} t_{\alpha\beta}~\d_{\lambda}
t_{\alpha\beta}$ analogously do not give the contribution to
nonderivative goldstone coupling, but the asymmetrical terms like
$\d_{\lambda} t_{\alpha\beta}~\d_{\alpha} t_{\lambda\beta}$ ~-
already can give. In fact these are the same terms in the kinetic
part of $\mathcal{L}_2$ that contribute to the spin
operators.
}.

From all possible field combinations, entering (\ref{aaa}) and linear
in $\omega_{\alpha\beta}$, only the term
\beal{lt}
\omega^{\alpha\beta} J_{\alpha\beta}~~~~\hbox{where}~~~~
J_{\alpha\beta} ~=~  ~\big(S^a_{\alpha}\d_{\beta}\varrho^a -
S^a_{\beta}\d_{\alpha}\varrho^a \big)
\big( S^a_{\mu}\d_{\mu}\varrho^a \big) ~+~
\d_{[\alpha} \omega_{..} n_{..} \d_. \omega_{\beta]}
\eea
gives the finite contribution in the soft limit
\footnote{
Note that terms (\ref{lt}) in $\mathcal{L}_2$, although linear in
$\omega_{\alpha\beta}$, does not contribute to the potential
$V(t_{\mu\nu})$ even after taking into account loop corrections.
}.
Expression (\ref{lt}), representing the soft interaction of
goldstone-gravitons with the rest of the fields (matter),
can be written in a more ``standard'' way
using (\ref{homega}):
$$
h_{\mu\nu} T^{\mu\nu}~,~~~~~~~~\hbox{where}~~~~
~T^{\mu\nu}~=~ 2 n^{\mu}_{\lambda} J^{\lambda\nu},~~~~
h_{\mu\nu}~\simeq~\omega_{\mu\beta} n_{\beta\nu} ~-~
    \omega_{\beta\nu} n_{\beta\mu}
$$

The kinetic terms for the goldstone fields $\omega_{\alpha\beta}$~
(the second line in (\ref{aaa})) can be expressed also
by means of $h_{\alpha\beta}$:
\bel{pf2}
\big( \d_{\gamma} \omega_{ab}
\d_{\sigma} \omega_{mn}\big)
\big( n_{\alpha\beta} n_{\delta\rho} \big)
 H^{\gamma a b~ \cdots~  \rho}~~~=\Rightarrow~~
\big( \d_{\gamma} h_{ab} \d_{\sigma} h_{mn} \big)~
\tilde{H}^{\gamma a b \sigma m n}~,
\ee
where the symmetric ``metric'' fluctuations $h_{\alpha\beta}$
are given by (\ref{homega}) and $\tilde{H}^{\gamma a b \sigma m n}$
in low energy limit is a constant tensor constructed
from $n_{\alpha\beta}$ and $\eta_{\alpha\beta}$.

If LI is restored at long distances then expression (\ref{pf2})
should reduce to sum  of the Pauli-Fierz Lagrangian
 \beal{pf}
   L_{pf} ~=~ \d_{\alpha} h_{\mu\nu} \d^{\mu}h^{\nu\alpha} -
   \frac{1}{2} \d_{\alpha} h_{\mu\nu}  \d^{\alpha} h^{\mu\nu}
   - f^{\alpha} \d_{\alpha}h  +
   \frac{1}{2}  \d^{\mu} h \d_{\mu} h~~,  \\
   \hbox{where}~~~~~~~  h = \eta^{\mu\nu} h_{\mu\nu},~~~~~~
   f_{\mu} = \d^{\nu} h_{\mu\nu}~~~~~~~~~~~~  \nn
 \eea
 and terms $\tilde{L}$, depending from $n_{\mu\nu}$ in such a form
that they can be considered as a gauge fixing terms (depending only
on $s_i$ and their derivatives).
 This non LI gauge can be defined by a condition which implies that
$h_{\mu\nu}$ can be represented in the form $h_{\mu\nu}~=~
\omega_{\mu\beta} n_{\beta\nu} ~-~ \omega_{\beta\nu}
n_{\beta\mu}$. Hence now $\omega_{\nu\beta}$ are new variables in
(\ref{pf}), $h = 0$, and only first two terms in right hand side
of (\ref{pf}) remain.

The antisymmetric fields $\omega_{mn}$ contain 6 local quantities,
connected with the massless spin two fields. Two of them represent
the transverse spin-two massless particles (gravitons). The
remaining four degrees of freedom are not propagating. In the GR
the corresponding quantities correspond to a Coulomb (Newton)
field coupled to the energy of sources, and to the 3-vector
$\vec{g}$ with components $\sim g_{0i}/g_{00}$ connected with the
angular velocity ($\sim rot~\vec{g}$) of the local system~
\footnote{ For the similar case of vector condensate $n_{\mu}
=\langle t_{\mu} \rangle $ we have 3 goldstone objects, formed
from $n_{\mu}$ by local transformations $\omega_{mn}(x)$ which
rotate $n_{\mu}$. Two of them represent the transverse vector
particles (photons), and third is the classical Coulomb field.
}.

Note that after spontaneous breaking of LI we have in fact a
theory with two metrics -- one is the flat background Minkowski
metric $\eta_{\mu\nu}$ and the other is the dynamical
`curved~metric'~$t_{\mu\nu}(x)$.

It is remarkable, that the goldstone gravitons do not interact
with the Lorentz invariant contribution to energy momentum tensor
$T_{\mu\nu} \sim \eta_{\mu\nu}$ because local Lorentz rotations do
not change it ~(See \cite{KrTomb}). The conditions (\ref{gaucon})
reflect this fact.
 Therefore in such a systems on can not have the de-Sitter type
behavior and thus the cosmological inflation stage driven by
almost constant scalar fields becomes problematic.

\subsection{\bf Degenerate case}

It is instructive to consider briefly the simplified case when
the condensate $n_{\mu\nu}$ is maximally degenerate:
$\lambda_1 = \lambda_2 = \lambda_3  \neq  -\lambda_0$.
Then in a coordinate system where $n_{\mu\nu}$ is diagonal one
has
$n_{11}= n_{11}=n_{11}=\lambda_1$, $n_{00}=\lambda_0$,
and there emerge only tree Goldstone modes.
In the weak field limit these are the components
$h_{i0}=\omega_{i0}(\lambda_0-\lambda_1)$.

This means that not all  gravitational configurations
can be reproduced in this case,
but only the ones that correspond to the ansatz
\bel{vmetr}
d s^2 ~=~ d t^2 ~-~ 2 v_i ~dx^0 dx^i ~-~
(\delta_{i k} + v_i v_k) ~dx^i dx^k~,
\ee
where $v_k(x^i,x^0)$ are some functions of $x^i$ and $x^0$.
Let us consider briefly the case of weak field,  with
$v_i \simeq h_{i0} \ll 1$.
Then the ``electric'' components  of the curvature tensor
read as $R^{0i0k} \sim \d^0 (\d^i h^{0k}+\d^k h^{0i})$ while
the other components are small in a weak field limit.
\begin{itemize}
\item
The gravitational field of static mass $m$ follows in this case from
equations  $R^{00} \sim m_p^{-2}T^{00} ~~\Rightarrow~~
 \d^0 \d^i h^{0i} \sim m_p^{-2} m \delta^3(x)$,
 with a simple solution
$$
  \omega^{i0} ~\simeq~ h^{i0} ~\simeq~ \frac{m}{m_p^2} \frac{x^0 x^i}{|x|^3}
  ~~,~~~~~~~ R^{0i0k} \sim (m/m_p^2) \frac{1}{|x|^3}
   \Big( \delta^{ik} - 3\frac{x^i x^k}{|x|^2} \Big)
$$
which leads to the Newton law~
\footnote{The generalization to strong fields can also be found in
this gauge.  It corresponds to Schwarzschild solution in some
special synchronous system  connected with form (\ref{vmetr}), and
i.e. applicable up to distances  $\sim |x^0| \sim |x^i| \sim
\Lambda^{-1}$ where the goldstones can disappear due to the
melting of  the tensor condensate $n_{\mu\nu}$.
}.
The force acting on a test particle is
$f^k \sim \Gamma^k_{00} \sim \d^0 h^{k0} \sim x^k/|x|^3$.

\item
For a uniformly distributed energy sources $T^{00}(x) = \varrho_0$ the
equation $R^{00} \sim \d^0 \d^i h^{0i} \sim \varrho_0$ gives regular
solution $h^{i0} ~\sim~ \varrho_0  x^i (x^0 \pm c)$, with nonzero
components of curvature tensor $R^{0i0k} \sim \varrho_0/m_p^2$.~
This corresponds to a 3-space metric at small $x^i$
$$
 g_{ik} = \eta_{ik} + h_{0i} h_{0k}
~=~  \eta_{ik} + \frac{x^i x^k}{a^2(x^0)}~~,~~~
   a(x^0) \sim \frac{m_p^2}{(c \pm x^0)\varrho_0}
  \sim a_0 (1 \pm \frac{\varrho_0}{ m_p^{2}} a_0 x^0 +...)~,
$$
with the time varying ``cosmological'' scale factor  $a(x^0)$
\footnote{Comparing this behavior of $a(x^0)$ with the case of
the cosmological scale factor in the GR
$(\d^0 a/a)^2 \sim \varrho_0 m_p^{-2}$, one can try to fix
the Hubble parameter as $H \sim \sqrt{\varrho_0}/m_p$.
But this is the result of linearized approximation.
Evidently, all these solutions are applicable only
in limited intervals of $x^0$ when $h_{0i}$ are small.
}.
The other way to ``measure'' this scale variation can be found from
the equation for geodesic deviation, which in the case of two standing
test particles, separated by the distance $\delta x^i$, reduces to
$\d^0\d^0 \delta x^i \sim R^{0i0k} \delta x^k$. Because
$(\d^0\d^0 \delta x^i) /\delta x^i  = (\d^0\d^0 a)/a$, this
corresponds to the standard evolution equation
$(\d^0\d^0 a)/a \sim \varrho_0/m_p^2$ for homogenous cosmology.
\end{itemize}
\nin  The goldstone fluctuations of the degenerate light-like
condensates of the type (\ref{n2}) can be used in the same way to
describe the transverse gravitational waves.

\subsection{\bf Einstein equations}
Going to the general case of strong goldstone fields it is
interesting to find under what conditions the first derivative
term (\ref{secterm}), without massive fields $\varrho^{a}$, can be
reduced to the Einstein-Hilbert Lagrangian, with $t_{\mu\nu}(x)$
playing the role of metric. The answer to this question, without
entering into the deep details, can be formulated in a rather
simple and natural form.

The {\bf \underline{first condition}} is that tensor $\Gamma^{
\alpha \beta \gamma \delta \rho \sigma}$ must be expressed only
through the ``contravariant''~ $t^{\mu\nu}$, which should be
defined as  inverse to $t_{\mu\nu}$, i.e. through
$t_{\gamma\nu}t^{\gamma\mu} =\delta^{\mu}_{\nu}$.
Solving this equation, we have
 \bel{contravar}
t^{\alpha\beta} ~=~ \frac{1}{4!\, t} \epsilon^{\alpha\lambda\rho\gamma}
    \epsilon^{\beta\sigma\mu\nu}
    t_{\lambda\sigma}  t_{\rho\mu}  t_{\gamma\nu}  ~,
\ee
 where $t = \det [t_{\mu\nu}(x)]$.
After that we get the general expression for~(\ref{secterm})~:
 \beal{gamma}
\L_2 ~=~~\L_2^{eh} ~+~
 \tilde{\L}_2~~~,~~~~~~~~~~~~~~~~~~~~~~~~~~~~~~~~~~~~~~~ \nn    \\
\L_2^{(eh)} ~=~ t_{\alpha\beta , \gamma} t_{\delta\rho , \sigma}
~\big( c_1 t^{\alpha\delta}~t^{\beta\sigma}~t^{\gamma\rho}~+~ c_2
t^{\alpha\delta}~t^{\beta\rho}~t^{\gamma\sigma}~+~
c_3 t^{\alpha\gamma}~t^{\beta\rho}~t^{\delta\sigma}~+~ \\
+~c_4 t^{\alpha\gamma}~t^{\delta\rho}~t^{\beta\sigma}~+~  \nn c_5
t^{\alpha\beta}~t^{\delta\sigma}~t^{\gamma\rho}
~\big)~~~,~~~~~~~~~~~~~~~~~~~~~~~~
 \eea
 where $c_i$ are some  scalar functions from invariants formed out
of $t_{\mu\nu}$. All such invariants can be expressed as functions
from $s_i$, and they should be taken at extremal values as far as
we have excluded heavy fields $\rho_i$. In this way, all $c_i$ in
(\ref{gamma}) are constants.
 The part $\tilde{\L}_2$  contains  terms with the same structure
as in (\ref{gamma}) but in which some ``contravariant''
$t^{\mu\nu}$ used for index contraction are changed by
$t_{\mu\nu}$ or $\eta_{\mu\nu}$.

Then the first condition is that  $\tilde{\L}_2 = 0$, or it can be
represented as a function from 4 scalar combinations $s_i$ defined
in (\ref{invs}) and of their derivatives in $x^{\mu}$. Then $\tilde{\L}_2$ can
be considered as gauge fixing term which does not change the
dynamics of $t_{\mu\nu}$ fields.
 Note that this condition, although simply formulated, is
dynamically very restrictive and it would be strange if it is
fulfilled ``by itself'' without fine tuning for some parameters in
the $\psi= (B,\chi,\varphi)$ field system.

The {\bf \underline{second condition}} is that
in the ``weak field limit''
$$
 t_{\mu\nu}(x) ~=~ n_{\mu\nu} ~+~ h_{\mu\nu} (x)~,~~~~~~
      h_{\mu\nu} ~\rightarrow~ ~0
$$
the expression (\ref{gamma}) transforms into the Pauli-Fierz
Lagrangian (\ref{pf}) for massless spin-two particles in a
nonorthogonal coordinates with metric  $n_{\mu\nu}$. In fact such
a condition removes the spin zero massless modes, which are
present in general in~(\ref{gamma}). As it can be simply shown it
leads to relations for coefficients in (\ref{gamma}):
$$
 c_2 ~=~ -2 c_1,~~~~~~~ c_3~=~c_4~=~c_5~=~0~~.
$$
These two conditions fix the form of $\mathcal{L}_2$ up to overall
constant ($c_1 \sim \Lambda^2 ~\rightarrow~ m_p^2$) and after that
we obtain finally that
 \bel{elagr}
\L_2^{(eh)} ~=~ c_1~ t_{\alpha\beta ,\gamma}
~t_{\delta\sigma,\rho} ~\big(2
t^{\alpha\delta}t^{\beta\rho}t^{\gamma\sigma} ~-~
t^{\alpha\delta}t^{\beta\sigma}t^{\gamma\rho} \big) ~,
 \ee
where $t^{\alpha\beta}$ is given by (\ref{contravar}).
 Possibly this condition can be relaxed, because in fact what
we need here is to truncate the contributions from spin 0 and~1
under specific ``gauge chose''.

This expression (\ref{elagr}) precisely coincides with the
Hilbert-Einstein Lagrangian in gauges which fixed
value of $t(x)=\det [t_{\mu\nu}(x)]= const$.
On the other hand,
here we have just this case - the ``gauge'' of goldstones in
$t_{\mu\nu}(x)$ is fixed by the conditions
\bel{gaucon}
 Tr(t_{\mu\nu})=a_1,~~~ Tr(t~t)= a_2,~~~
 Tr(t^3)=a_3,~~~Tr(t^4)=a_4~~,
\ee
where constants $a_k$ coincide with roots of equations
$\d V/\d s_k = 0$ ~(See (\ref{roots})), connected to eigenvalues
of $t_{\mu\nu}$ and as a result the  ~$\det [t_{\mu\nu}(x)]=
\lambda^{(1)}\lambda^{(2)}\lambda^{(3)}\lambda^{(4)}= const$.
This is by itself very interesting gauge. In this case $\det
[t_{\mu\nu}]$ is also fixed by the fact that the
$t_{\mu\nu}$~-~goldstones correspond to the 4-volume preserving
fluctuations of metric. In other way, this explains   why the
goldstone gravitons does not interact with the terms corresponding
to cosmological constant.

The Hilbert-Einstein Lagrangian $\L_2$ in (\ref{elagr}) is
polynomial in $t_{\mu\nu}(x)$,  and in fact such a theory looks
like some special $\sigma$-model
\footnote{ The accidental scale-invariance of $\tilde{L}$ in
(\ref{elagr}) of type $x  \rightarrow a \, x$, $t_{\mu\nu}
\rightarrow a^{2/11} t_{\mu\nu}$ is probably fictitious because
gauge fixing conditions (\ref{gaucon}) are not invariant under
such a transformation.
}.
 It takes even more simple form, when we substitute in (\ref{elagr})
and (\ref{contravar})  $t_{\mu\nu}$ in form (\ref{omen})
and use $\Omega^{\mu}_{\nu}$ as a main variables, describing the
gravitational field.

The terms with higher then two powers of derivatives in
(\ref{lagran}) can lead to higher order in ``curvature'' terms in
$\L$, and near to the ``Planck'' scale where $\d_{\mu} \sim
\Lambda$ all these terms can be of same order as $\L_2$

Here again as for $\L_2$ we come two type of terms. Terms that we
construct by using only the covariant $t^{\mu\nu}$ for contraction
of indices and other terms - in which $\eta^{\mu\nu}$ and
$t_{\mu\nu}$ are also used. Then from the first we come to
expressions containing invariants constructed only from higher
powers of the curvature tensor and taken in gauge (\ref{gaucon})~
\footnote{
In fact the same situation takes place for the usual
geometrical gravity, where the ``natural'' general form of the
Lagrangian is $\sum_n c_n R_n$ where $c_n \sim m_p^{2n}$ and the
invariants $R_n \sim (R_{abcd})^n$ are constructed from n-th
powers of the curvature tensor.
}.
The other terms are different,  and perhaps they can be
interpreted as the gauge fixing terms for the ``invariant'' part
of the Lagrangian. In this case we have the Lorentz invariant
description at all scales.

The last possibility is attractive but unfortunately we do not see
clear reasons for its realization.  Less restrictive is the possibility
that the coefficients entering in such an expansion can be scale
dependent, and it is not excluded that at distances $|x_i-x_j| \gg
\Lambda^{-1}$ only the invariant part survives  and the
``additional'' terms becomes comparatively small, for example like
~$\sim 1/(|x_i-x_j|\Lambda)^2$ or turn into ``gauge fixing''
expressions.

\section{\bf  Tensor condensates in curved space and bigravity theory}

Here we briefly consider what modifications can be expected if we
repeat the constructions of previous sections with slowly varying
condensate $\langle \tau_{\mu\nu}(x) \rangle $ and its
fluctuations in the external curved background $g_{\mu\nu}(x)$,
instead of flat Minkowski metric  $\eta_{\mu\nu}$ without back
reaction on metric $g_{\mu\nu}(x)$ from the system of
$\psi=(B,\chi,\varphi)$ fields.

For  small curvatures,
$|R_{\mu\nu\lambda\sigma}(x)| \ll \Lambda^2$,
the local dynamics which leads to the condensation
of some composite operators in field system $\psi=(B,\chi,\varphi)$
remains approximately the same as in a flat case.
Therefore, the mean value
$$
\langle \tau_{\mu\nu}(x) \rangle ~=~  n_{\mu\nu}(x)~~,
$$
should be a covariantly constant tensor which in the local
Lorentz frame for every point $x$ reduces to
approximately the same $n_{\mu\nu}$
as before, with a possible corrections of order
$|R_{\mu\nu\lambda\sigma}(x)|/\Lambda^2$.

If we now consider the geometrical gravity as a dynamical
field also, then on should expect that in a weak field limit
the propagation of $t_{\mu\nu} = n_{\mu\nu} + h_{\mu\nu}$ and
$g_{\mu\nu} = \eta_{\mu\nu} + \beta_{\mu\nu}$ will be independent.
And the form of effective  $\L(h_{\mu\nu},\beta_{\mu\nu})$ - in first
approximation is given by a sum of two Pauli-Fierz Lagrangians for
$h_{\mu\nu}$ and $\beta_{\mu\nu}$~ (if propagation of
$\beta_{\alpha\beta}$ is Lorentz-covariant).
The interaction of these fields with other  massive particles will
be also independent in this approximation.
If in addition the goldstone gravitons interact in the long range limit
universally (i.e. with the energy momentum tensor),
then their actions will be indistinguishable.

It seems that the dynamical mixing between $t_{\mu\nu}$ and
$g_{\mu\nu}$ is impossible (as between massless particles), so at
long distances there remain two massless particles, which  act
in a rather similar way.
The presence of the second (true) `gravitational field'
can be manifested only in possible small nonuniversality
of $t_{\mu\nu}$ interactions.

When both fields $t_{\mu\nu}$ and $g_{\mu\nu}$ are strong
(it is when scales $\Lambda_s$  and $\Lambda$ are also close)
then their interaction can be highly asymmetrical and complicated.

But if a coupling constants in the `fundamental lagrangian'
$L(\chi,B,\varphi)$ are such that the difference between
$\Lambda$ and the geometrical Planck(string) scale $\Lambda_s$ is
very big then the quantum fluctuations of geometry near the scale
$\Lambda$ would be small, and then the geometrical and goldstone
gravity will in fact decouple.
 In this case  only the composite gravity ($t_{\mu\nu}$) is essential
at the macroscopic distances.

The geometrical gravity can again start to contribute only at
maximal cosmological scales (See Section 6) du to the interaction
with various contribution to $\langle T_{\mu\nu} \rangle \sim
g_{\mu\nu}$ imitating the cosmological constant~
\footnote{It is not excluded that is possible to remove one of
the massless particles in ($t_{\mu\nu}, g_{\mu\nu}$) system by
some form of Higgs effect for tensor particles. For this we
additionally need massless vector and scalar Goldstone particles
interacting in a special way with $t_{\mu\nu}$. May be the
spontaneous breaking of LI is in some ``virtual'' form realized
also in QCD like theories, so that the corresponding Higgs effect
makes the tensor and vector goldstones massive  - and the remnants
of it are the massive vector and tensor resonances, whose
interactions reminds the gauge coupling}.

\section{\bf  The Goldstone  graviton coupling to other particles:
universality problem}

In the geometrical (string) gravity the gravitons are universally
coupled to all particles and this is probably one of the main
experimentally established property of the gravitation. The
general relativity `explains' this fact in a simple and natural
way. For the $t_{\mu\nu}$ gravitons the situation is more
complicated.

Note that in fact we ``do not need'' the universality of
$t_{\mu\nu}$ (gravitons) interactions at all scales, but only in
the long-distance limit, where we know (from experiment) that it
is fulfilled with an extreme precision.

Since goldstones $t_{\mu\nu}$ are not geometrical objects, the
universality of their coupling with other fields are not automatic
at all scales and must be probably tuned to reach the agreement
with the experiment. At  the same time, although the local
Lorentz-invariance is spontaneously broken, the translation
invariance remains. Usually just this invariance leads, after
localization, to diffeomorphism-invariance and universality of
graviton interactions. Therefore, one can hope that at the
distances much larger than $\sim \Lambda^{-1}$, at which the local
Lorentz-invariance is broken, the approximate diffeomorphism
invariance remains, and as a result we come to the universality of
the long-range goldstone graviton mediated interactions.

But evidently  at the scales close to $\sim \Lambda^{-1}$ one must
expect big nonuniversalities. This can be directly seen in a
models of the following type.

Suppose that in addition to the sector of $\psi$-particles
that we considered before,
$\psi = (B\,\chi,\varphi)$, and from which the Goldstone
modes are formed,  there exist also a sector of $s$ particles,
only weekly coupled to particles of $\psi$ sector by some
mediator (gauge) particles $\gamma$ with small coupling~$e$.
This can be, for example, the sector of
standard model or some of its generalization, and
$\gamma$ - some gauge fields coupled to both sectors.

\begin{figure}[h]
\centerline{\epsfxsize=4.9in\epsfbox{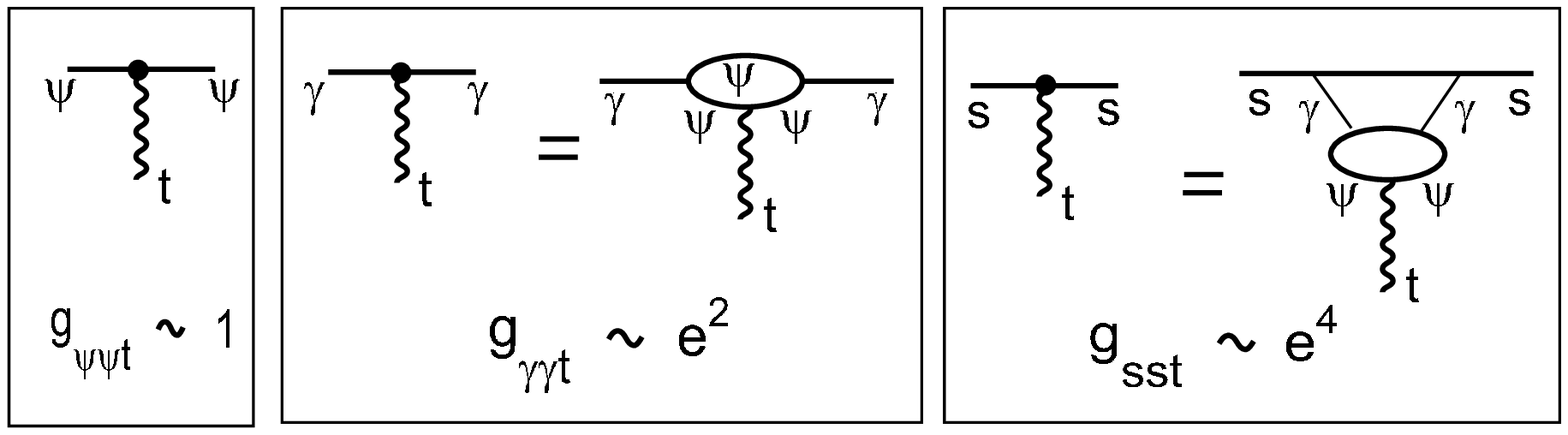}}
\caption{ \small
Simple model with with weekly coupled sectors $(\psi)$  and ($s$) and
mediator particles ($\gamma$).
Vertexes, the main diagrams, and their orders in $e^2$ are shown
\label{Figure1} }
\end{figure}

Then we have tree classes of values of couplings
~$(\sim 1,~ \sim e^2,~ \sim e^4)$ for goldstone-graviton vertexes
$(t\psi\psi,~~t\gamma\gamma,~~tss)$, corresponding to diagrams
shown in Fig. 1.
Because~$e^2$ can be ``chosen'' arbitrary small on the scale $\Lambda$,
on can have very big nonuniversality of $t_{\mu\nu}$ couplings
at scale $\Lambda$.
And even if the general universalization mechanism exists,
here it can equalize couplings only at very large distances
\footnote{ Let us remark, however, that we know very little (from
a direct experiment) whether the gravitational interaction of some
weekly interacting particles (for example dark matter) is exactly
universal}.

The same nonuniversality can be directly seen for the system
described by the Lagrangian $\L_2$. The interaction of
$\omega_{ab}$ goldstones with other particles is contained in
terms entering the equation of motion
 \bel{eillag}
\frac{\delta~\mathcal{L}_2(t_{\mu\nu})}{\delta\omega_{ab}(x)} ~=~
\frac{\d\mathcal{L}_2}{\d \omega_{ab}} ~-~
\d_{\mu}\frac{\delta~\mathcal{L}_2}{\delta(\d_{\mu}
\omega_{ab})does not} ~+~ ...
 \ee
In the long range limit only the
$\d \mathcal{L}_2/\d \omega_{ab}$ therm in (\ref{eillag})
contributes to the interaction-current,  connecting $\omega_{ab}$
with other fields~:
 \bel{omcur} \frac{\d \mathcal{L}_2}{ \d
\omega_{ab} } \simeq \big( \Gamma^{......}
Q_{\mu\nu}^{ab~\gamma\sigma}\big) \big( \d_{\gamma}
\t_{\alpha\beta} \d_{\sigma} \t_{\delta\rho} \big) ~=~
\Upsilon^{\mu\nu}_{ab} \big( \d_{\mu}\varrho^n ~\d_{\nu} \varrho^n
\big)~,
 \ee
where $\Upsilon^{\mu\nu}_{ab}$ some constant tensors.
The structure of  $\omega_{ab}$ interactions, coming from the term
$\d \mathcal{L}_2/\d \omega_{ab}$ in (\ref{eillag}), evidently
remains the interaction of gravitons in general relativity, coming
from analogous terms $T_{\mu\nu} = \d L/\d g_{\mu\nu}$. Note that
in the week field limit $\omega_{ab}$ and $h_{\mu\nu}$ are
linearly related as in (\ref{homega}).

The fundamental fields and physical particles do not enter
in $\varrho^m$ symmetrically on the scale $\Lambda$.
Therefore, as follows from (\ref{omcur}), there is no universality
in the goldstone-graviton coupling at such a momenta.
On much larger distances $x_{i j} \gg \Lambda^{-1}$ operators
$\d \mathcal{L}_2/\d \omega_{ab}$ in (\ref{omcur}) are renormalized,
and so is not excluded that  for $t$-gravitons with momenta
$k \rightarrow 0$ the universality is reached.

There exist rather simple general argument \cite{wein} that
in Lorentz-invariant case the massless spin two particles must
be coupled universally to all particles in the limit
$k \rightarrow 0$.
This construction can be generalized to the case with spontaneous
breaking of LI as follows.

\begin{figure}[h]
\centerline{\epsfxsize=1.9in\epsfbox{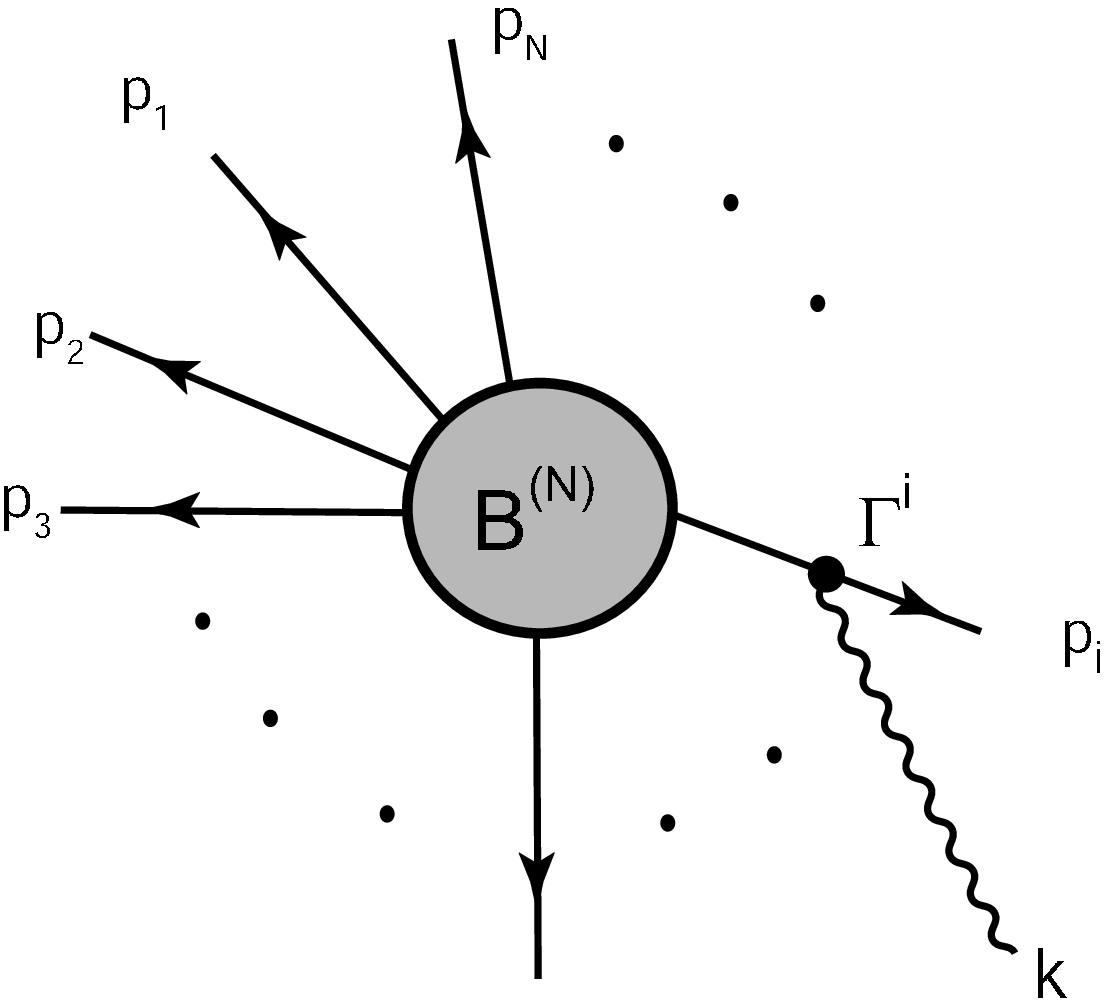}}
\caption{
\small The general $N$ particle amplitude for a soft graviton emission
with a momentum $k$.
\label{Figure2} }
\end{figure}

Consider some general N-particle amplitude with soft emission of
$t_{\mu\nu}$-graviton with momenta $k\rightarrow 0$.
 As usual, in this limit the main contribution comes from diagrams
with the graviton emission from all external lines (See Fig.2),
and has the form:
$$
 A_{\mu\nu}^{(N)}(p_1,...,p_N,~k)~\simeq~ B^{(N)}
       (p_1,...,p_N) ~\sum_{i=1}^N
    \frac{\Gamma_{\mu\nu}^i(p_i)}{(2 p_{i\alpha}
     ~k_{\beta} ~\G^i_{\alpha\beta})}~~,
$$

where we supposed that propagators of  $p_i$-particles near mass
shell have the form $2\big(p_{i\alpha}  ~k_{\beta}
~\G_{\alpha\beta} - m^2_i\big)^{-1}$, and the tensors
$\G^i_{\alpha\beta}$ can depend from  $n_{\mu\nu}$.
 The vertexes  $\Gamma^i_{\mu\nu}$ for a on mass shell graviton
emission by particle of type~(i) can also contain all possible
combinations of
$p_{\mu}$ and $n_{\mu\nu}$
 \beal{gaver}
\Gamma_{\mu\nu}^i(p) ~=~ C^i p_{\mu}p_{\nu} ~+~
(H^i_1)_{\mu\nu} ~+~
p_{\mu}~(H^i_2)_{\nu\lambda}~p_{\lambda} ~+~ \nn \\
+~p_{\lambda}~(H^i_2)_{\lambda\mu}~p_{\nu} ~+~
(H^i_3)_{\lambda\mu} ~p_{\lambda}~(H^i_3)_{\nu\beta} ~p_{\beta}~~,
 \eea
where the tensors
\bea
(H^i_k)_{\mu\nu} ~=~ (C_k^i n + \hat{C}_k^i nn + \tilde{C}_k^i nnn +
  \check{C}_k^i nnnn )_{\mu\nu}~,~~~i,k=1,2,3
\eea
and $C_k^i,~\hat{C}_k^i,...$  - some invariant
functions depending at $k=0$ only from particles type.
In  the usual case without LI breaking we have in (\ref{gaver})
only the first term  $p_{\mu}p_{\nu}$ in graviton particle coupling.
Du to masslessness of $t_{\mu\nu}$-gravitons the amplitudes
$A_{\mu\nu}^{(N)}$ should be transverse
\footnote{This point can look not a certain because LI is already broken.
It corresponds to removing of scalar part of $t_{\mu\nu}$ with respect
to $\eta_{\mu\nu}$ .}
at all N and particles momenta~$p^i$~:
\beal{trcon}
2 k_{\mu} A_{\mu\nu}^{(N)} ~=~ B(p_1,...,p_N)~\sum_{i=1}^N~ \Big(~
 C^{i}p^i_{\nu} ~+~
\big( H^i_2 p^i  \big)_{\nu} ~+~ \nn \\
 + ~\frac{1}{(\G^i_{\alpha\beta}k_{\alpha} p_{i\beta})}
 ~\big(~   (H^i_1 k)_{\nu}  ~+~
(k H_2 p^i)_{\nu} ~+~
(p^i H_3)_{\nu} (k p^i)   ~\big)  ~\Big) ~=~ 0~~~.
\eea
The only possibility that this conditions can be fulfilled identically
at all N and $p^i$ is when equation (\ref{trcon}) is reduced
to energy-momentum conservation $\sum_i p^i_{\mu}=0$.
And the momentum conservation take place because the translation invariance
is unbroken by tensor condensate.
This impose conditions on coefficients entering equation (\ref{trcon})
which has the general solution of the form:
$$
(H_1)^i_{\mu\nu} =0~;~~~
(H_2)^i_{\mu\nu} = a_i ~\G^i_{\mu\nu}~;~~
(H_3)^i_{\mu\nu} =~b_i ~\G^i_{\mu\nu}=~b_i ~\eta_{\mu\nu}~;
$$
$$
C^i + 2a_i +(b_i)^2 =~ C .
$$
And so we end with the universal (for $k \rightarrow 0$) vertexes
for the soft  $t_{\mu\nu}$-graviton interactions with particles
\bel{gacor}
\Gamma_{\mu\nu}^i(p,k) ~=~ C p_{\mu}p_{\nu} ~+~
\frac{k^2}{\Lambda^2}~\hat{\Gamma}_{\mu\nu}^i(p) ~+~ ...~,
\ee
where $\hat{\Gamma}_{\mu\nu}^i(p)$ are nonuniversal and can
depend on $n_{\mu\nu}$.

The details of mechanism, which can lead to the behavior of the type
(\ref{gacor})) are not quit clear,  but it can be somehow
connected to a different scale behavior of operators $\sim p_{\mu}
p_{\nu}$ and the dependent on $n_{\mu\nu}$ contributions to
vertices $\Gamma_{\mu\nu}$.

If we suppose that (\ref{gacor}) take place, the corrections to
`universal' vertexes (which $\sim T_{\mu\nu}$), from various
nonuniversal operators in $\tau_{\mu\nu}$ can decrease with scale
$\omega$ as $(\omega/\Lambda)^n$, where $n \ge 1$.
 For $\omega < 10^{-3} \div 10^{-5}~ eV$, where we already have
experimental, data confirming the universality of gravity,
the factor $(\omega/\Lambda)$ is $< 10^{-31}$.
This can be compared with the best experimental limits on
the universality of $|1 - m_{grav}/m_{inert}| < 10^{-15}$,
coming from last Etvesh like measurements
\footnote{
This comparison also shows that the minimal value on which on
can now lower the value of $\Lambda$ by various brane-world
mechanisms is $\ge 10^3 GeV$.}.

From a more general point of view, when on looks on the
possibility of restoration of maximal symmetry on long distances,
many example of this known.
 But  it is complicated to formulate the universal criteria for this.
The simplest examples come from various lattice calculations (and
experiments) - here not only isotropy but even translation
invariance is broken on the scale of lattice spacing $a$. But at
distances $x \gg a$ the fast isotropization take place, and the
direction dependent contributions decrease as powers of $(a/x)$.~~
 On should also mention the calculations in
\cite{nil2},\cite{nil2}, where the isotropization  in
renormalization flow to large distances is observed. But at the
same time we know that there exist also such a media, whose
macroscopic anisotropy reflects its the microscopic asymmetry.

From various of examples, we know, that the main reason for such
an isotropization is connected with the fact that the quantities
responsible for the spacial symmetry breaking are represented by a
\underline{dimensional} quantities (like lattice spacing or tensor
$\Lambda^4 n_{\mu\nu}$). Therefor the operators, in which their
enter, have different scale dependence, in comparison with more
symmetric operator contribution, representing the same physical
quantity.

Note, that  the different approach to the universality of
interactions of nonscalar goldstone particles was proposed
\cite{jon}~. In \cite{jon} the primary condition is that the
non-Lorentz invariant terms in the effective Lagrangian can affect
only the ``gauge-dependent'' components of Green functions, so
that they should not enter (at all frequencies) into measurable
quantities.
 Is unclear for us if is possible in dynamical theory of the
type we consider, i.e. in the situation when there exist a real
``oriented'' condensate, which in principle can enter in various
amplitudes.

\section{\bf  Supersymmetric generalization}

 It is interesting to generalize the above construction of
spontaneous breaking of LI and of corresponding
graviton-goldstones to the supersymmetric case.
 This, in particular, can also be essential for the cosmological
mechanisms and various unification models. In bigravity models by
changing the relation between the geometrical an goldstone scales
on can try to regulate the value of the long-scale cosmological
constant.
 Besides, this can be natural for  bigravity models, when on a
small distances we have the string description, witch usually
leads to a supersymmetric field theory on a larger distances.
 If  the ``basic'' field theory $(B,\chi,\varphi)$ is
supersymmetric then on the larger scale, where the $n_{\mu\nu}$
condensate, we have two main possibilities:
\begin{itemize}
\item
the first case - when the supersymmetry is also broken at
the scale $\Lambda$;
\item
the other case -  the supersymmetry is
not broken by the mechanisms which generate the $<\tau_{\mu\nu}>$
condensate and is broken by the other mechanism at much larger
distances.
\end{itemize}

The difference between these versions is evidently reflected in
the behavior of $<~T_{\mu\nu}~>$.
In a first case we can have nonzero mean $T_{\mu\nu} \sim
\Lambda^4$ in (\ref{tmn}),~ and in the case of non broken
supersymmetry we have $<T_{\mu\nu}> = 0$ at $\Lambda$ scale and at
some lager distances.
In (\ref{tmn}) this looks as an additional relation between
coefficients. The first possibility is more general.
And in this Section we briefly consider the second case, to define
conditions when it can take place.

So, as above, we suppose that there  exist special sector, now
supersymmetric, with becomes strongly coupled on some scale
$\Lambda$.
We assume that the dynamics of this sector is described by the
effective renormalizable theory - otherwise it is complicated to
have the grow of couplings with distance.
Its lagrangian $L(V^i,\phi^i)$ can contain only vector gauge
superfields $V^i$ in adjoint representation of some nonabelian
group (say $SU(N)$)~,~ and also a number of chiral superfields
$\phi^i$ in multiplets of the same group.
These fields also interact with a sector of ``our'' supersymmetric
matter (using some mediators).
 Such Lagrangian is fixed up to values of constants, and  something
is known about the strong coupling behavior of such a theories.
 And we want to know if it is possible such a situation, when some
vector and (or) tensor quantities have nonzero vacuum expectation
values, but at the same time the supersymmetry is not
spontaneously broken, and how natural is such a behavior. Here we
use the simplest approach to this question.

Begin from the supersymmetric generalization of tensor operator
$\tau_{\mu\nu}$ decomposition (\ref{oper}) in terms of the
fundamental renormalizable superfields $V^i , \phi^i$. For this we
firstly include $\tau_{\mu\nu}$ in some adequate superfield as its
lowest component. If we simply act on $\tau_{\mu\nu}$ by the
global supersymmetry transformation
 \bel{suptran}
H_{\mu\nu}(x,\theta,\bar{\theta}) ~=~ \exp (\xi \bar{Q} +
        Q \bar{\xi} )~ \tau_{\mu\nu}(x)
\ee
with parameters $\xi$ and generators $Q$, we also generate the other
components of this superfield. Such a superfield is in general
reducible - sum of real plus chiral and antichiral.
We suppose that only chiral (plus antichiral) component
is present
\bel{h1}
H_{\mu\nu} ~=~ \tau_{\mu\nu}~+~
   \theta \chi_{\mu\nu}~+~ \theta\theta f_{\mu\nu}
\ee
Such a chose of multiplet, in which we inserted $\tau_{\mu\nu}$,
essentially restricts the model and allows not to break LI on
condensate scale \footnote{Note that if $\tau_{\mu\nu}$ is chosen
as component of a real superfield, than probably we will always
break supersymmetry together with LI by condensing
$\tau_{\mu\nu}$.}.
We can decompose
\beal{hvv}
H_{\mu\nu} ~=~ U^a_{\mu} U^a_{\nu} ~=~ s^a_{\mu} s^a_{\nu} ~+~
  \theta~(s_{\mu}^a \chi^a_{\nu} + \chi_{\mu}^a s^a_{\nu}) ~+~ \nn \\
     +~ \theta\theta~(s_{\mu}^a \sigma^a_{\nu} + \sigma_{\mu}^a s^a_{\nu}
     + \chi_{\mu} \chi^a_{\nu})~,
\eea
so that the lowest component of chiral superfield
$$
U^a_{\mu} ~=~ s_{\mu}^a + \theta \chi_{\mu}^a +
         \theta\theta \sigma_{\mu}^a
$$
coincides with vierbein eigenvectors $s^a_{\mu}$ ~(see (\ref{ev})),
in with we also included factors $\sqrt{\lambda^a}$ from eigenvalues.

In terms of `fundamental' superfields $V^i$ and $\phi^i$ the operator
decomposition of $\hat{H}_{\mu\nu}$ ~~(the analog of (\ref{oper})) can
look as
\bel{h2}
\hat{H}_{\mu\nu} ~=~ P_1(\phi)~\d_{\mu}\phi^i ~\d_{\nu}\phi^i ~+~
   P_2(\phi)~ W^i_{\alpha}\sigma_{\mu}^{\alpha\dot{\beta}} \d_{\nu}
    W^i_{\dot{\beta}} ~+~...~~,
\ee
where $W^i_{\alpha} = \bar{D}\bar{D}D_{\alpha} V^i$ - is supergauge
strengths, $P_i$ some real functions from chiral fields $\phi^i$
and summation over indexes of gauge group G is assumed.

After we integrate over fields $V^i$ and $\phi^i$ ~(at fixed
$H_{\mu\nu}(x,\theta))$ we come to  (as in~(\ref{genfun1}))
the supersymmetric effective lagrangian $\L(H_{\mu\nu})$
depending on $H_{\mu\nu}(x,\theta)$.
We write it in a  form of symbolic decomposition in powers of
superfield $H_{\mu\nu}$ and its derivatives in coordinates:
\bel{lh}
\L(H_{\mu\nu}) ~=~ F^{(0)}(H)_{\2t} ~+~
       F^{(1)}(H \cdot H^{+})_{\4t}  ~+~
  \d_{\alpha} \d_{\beta} F^{(2)}_{\alpha\beta}(H~H^{+})_{\4t}
  ~+~ ...
\ee
with some functions $F^{(i)}$, defined by the dynamics on scale $\Lambda$.
The first terms contribute to potential and higher describe the
propagation of degrees of freedom contained in $H_{\mu\nu}$.
The simple example of terms in (\ref{lh}) is given by
$$
\L(H_{\mu\nu}) ~=~ F^{(0)}(H)_{\2t} ~+~(H~H^{+})_{\4t} ~+~ ...
$$
The corresponding potential energy will have the form
\bel{suptl}
V(t_{\mu\nu})~=~ f_{\alpha\beta}(t) f_{\alpha\beta}^{+}(t)~~,
\ee
where
 \bel{fterm}
f_{\alpha\beta}^+(t) ~=~ a_1 t_{\alpha\beta} ~+~
a_2 t_{\alpha\gamma} t_{\gamma\beta} ~+~
a_3 t_{\alpha\gamma} t_{\gamma\delta} t_{\delta\beta}~+~ ...
 \ee
is the highest component of $H^+_{\alpha\beta}$, and the
coefficients $a_i$ correspond to expansion of $F^{(0)}$ in powers
of $H$. Expression(\ref{suptl}) is the special case of potential
(\ref{orpot}) which now it is positive definite.
This follows because we supposed that in every point
$t_{\alpha\beta}$ can be reduced to diagonal form by local
Lorentz-transformation.
This leads to diagonal form of $f_{\alpha\beta}$ and to positivity
of right hand side of (\ref{suptl}).
Therefore  ~$t_{\alpha\beta} = 0$~ is always the solution,
corresponding to a minimum of potential with $V(0)=0$.~ In this
case LI is not broken spontaneously.

There can also exist other minima in (\ref{suptl}) with
$t_{\alpha\beta} = n_{\alpha\beta} \neq 0$~,~
and $V(n_{\alpha\beta})=0$ ~-~ at least for some regions
of parameters $a_i$.
In diagonal basis for $f_{\alpha\beta}$ we become for potential
$$
 V ~=~ \sum_{i=0}^3 ~(f(t_{ii}))^2~~,~~~~ f(t)~=~
 \sum_{m=1}^{\infty} ~a_m t^m ~~.
$$
And because the equation $f(t)=0$ can have many nontrivial zeros
without fine tuning of $a_m$,  such zeros will generate minima
of $V(t_{\alpha\beta})$ with the same $V(n_{\alpha\beta})=0$.
And therefore supersymmetry is not yet broken in these minima,
but the LI is broken
\footnote{
In a more general case (\ref{lh}) the potential will have similar
to (\ref{suptl}) form
$
V(t_{\mu\nu})~=~ f_{\alpha\beta}(t) f_{\alpha\beta}^{+}(t)~
/F(t t^+),
$
where $F(z) \sim \d  F^{(1)}(z) /\d z $.
Here we come to similar conclusions, although the region of allowed
parameters, for not to brake supersymmetry, can be more narrow.
},
and these minima are stable
\footnote{ For this case we can have the domain structure of vacuum
in with regions with different Lorentz  structures
$n_{\mu\nu}$ are separated with massive supersymmetric walls.
At large distances the supersymmetry expected to be broken,
and domain structure becomes unstable.}.

At scales where the supersymmetry is  broken the energy of these
vacuums can become different -  of the order of supersymmetry
breaking scale~$\Lambda^4_{ss}$. Which of them will be lower,
depend from details of dynamics  - this was already discussed in
previous sections.

The oscillations of $t_{\alpha\gamma}$ corresponding to the local
Lorentz rotations give the goldstone gravitons - for long
wavelength they almost does not change the potential
$V(t_{\mu\nu})$. But here the the local supersymmetry
transformations (\ref{suptran}) with Majorana spinor parameters
$\xi(x)$ also do not change the energy in the long wavelength
limit and correspond to massless goldstone fermions - gravitinos.
 These $\xi(x)$ give the transverse components of such
goldstone-gravitino in a specific gauge, defined by the condition
that in every point gravitino field $\chi_{\mu}(x)$ can be
transformed into the same constant eigenvectors $ S^a_{\mu}$ by
the local supersymmetry transformation with parameters $\xi(x)$.
That is goldstone-gravitino can be represented in the form
$$
\chi_{\mu} (x) ~\sim~ \Big( \exp \big( i Q \xi(x) \big)
 -1 \Big) ~\sigma_a S_{\mu}^a ~.~
$$
in such gauge.

After substitution of (\ref{h1}) in (\ref{lh}) we come to some
effective constrained supergravity lagrangian, where gravitinos
are described by spinor $\xi(x)$ and gravitons - by the
antisymmetric matrixes $\omega_{\mu\nu}(x)$ corresponding to local
Lorentz transformations .

Below the supersymmetry breaking scale these gravitinos are
already not a purely goldstone particles and become a mass,
depending from mechanism by which the supersymmetry breaking takes
place.

\section{\bf  Cosmology}

If goldstone-gravity lagrangian approximately coincides with
Einstein lagrangian as in (\ref{elagr}), then probably all main
stages of standard Friedman-Robertson-Walker (FRW) cosmological
model can ran here without additional tuning.

There is even some advantage - at temperatures $> \Lambda$
the composite gravity disappears (because the $<t_{\mu\nu}>$
condensate melts), and it gives additional possibilities
to avoid the cosmological singularity.

The important property of goldstone gravitons is that they do not
interact with contributions to $<T_{\mu\nu}> \sim \eta_{\mu\nu}$.
 This is so because this expectation value is not changed by the
local Lorentz transformations $\Omega_{\mu}^{\lambda}(x)$
which ``contain'' the graviton-goldstone degrees of freedom.
In particular, this means that the inflation stages can not appear
in such a purely goldstone gravity from large coherent
fluctuations of some effective scalar field .

In the more interesting (bigravity) case
\footnote{  Number of models was proposed which include
many   ``metric''  and which look geometrically unusual
but du not  contradicting to experiment.
See for example \cite{dako}, \cite{BCNP}
}
when the geometrical gravity is also present,  only the
geometrical component of metric can  ``feel'' the
$\langle T_{\mu\nu}\rangle ~\sim~\eta_{\mu\nu}$ component.
Then the fundamental geometrical planck scale $M_P$ can
be chosen so that $M_P \gg \Lambda$. And this way on can
``naturally'' make the observable cosmological constant small.

 In the next sub-sections we briefly consider main features
of the two type scenarios.

\begin{itemize}
\item

In first there is no fundamental geometrical gravity - only
of the goldstone type.

\item
In the second case we have special bigravity~: there are both -
geometrical and goldstone tensor fields with scales adjusted such
that $M_P \gg \Lambda$
\footnote{ One can also mention the even more sophisticated
possibility, when the field theory model with only the
goldstone-composite gravity (for some specially adjusted
$(B\,\chi,\varphi)$ fields) is dual to string systems (in a
special background), with contain long-range ``geometrical''
gravity. Then on can even expect that the cosmology with a pure
goldstone-gravity should give approximately the same picture as a
some solution of standard geometrical cosmology.}.

\end{itemize}

Below we also suppose that all possible macroscopic effects coming
from a possible  breaking of LI on scale $\Lambda$,  du to existence of
$n_{\mu\nu}$~,~ can be made small - less than seen in current experiments.

\subsection{Cosmology with only goldstone gravity}

Here the most natural cosmological scenario corresponds to FRW
geometry with spatially flat infinite metric with the ``standard''
stages with scale factor $a(t) \sim |t|^{\alpha}~~\alpha =
1/3,~2/3$~. Now this is applicable in time intervals $-\infty < t
<  -t_0 ,~~ t_0 < t < \infty$.
The value of $t_0$ is defined such that at $-t_0 < t< t_0$ the
temperature of media is greater then $\Lambda$, so that the
$<\tau_{\mu\nu}>$ condensate evaporates and there is no
gravitational interaction in this time interval.
Therefore the compressed media at $-t_0 = t$ enters in the
inertial stage of compression, and as a result the compression
stops at some maximal density of order $\Lambda^4$ (we can chose
time t=0 for this moment). After that the media begins to expand,
initially without deceleration, but the temperature is falling.
\begin{figure}[h]
\centerline{\epsfxsize=1.9in\epsfbox{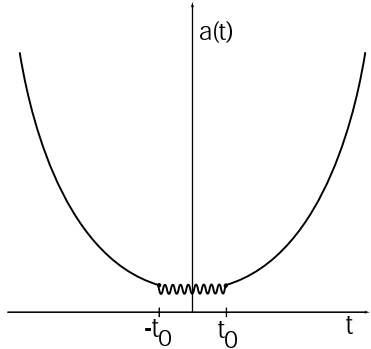}}
\caption{
\small Scale factor variation with time. In the interval from
$-t_0$ to  $t_0$ the Goldstone-gravity is switched off.
\label{Figure3} }
\end{figure}
When temperature reaches some critical value of order $\Lambda$
the $<\tau_{\mu\nu}>$ condensate appear and the goldstone gravity
is switched on again and the system enters at the
$t_0 < t < \infty$  an FRW stage
\footnote{In this model the initial conditions
for $<\tau_{\mu\nu}>$ and the matter fields at $t=-\infty$ can be
prepared in some complicated nonisotropic and nonhomogeneous form.
But the configuration of newly created $<\tau_{\mu\nu}>$
condensate at $t > t_0$ will be without any long range order, and
therefore at $t \gg t_0$ we come to the meanly flat space metric,
and the small scale inhomogeneities will decay to particles or
gravitational waves.}.

In this scenario some contemporary experimental data about the
character of universe expansion, which are most often interpreted
using small cosmological constant, can only be explained  by some
different mechanisms - possibly coming from some  universal
corrections.

Note that in such a model we do not meet the naturalness-type
problems with flatness, isotropy etc., which are usually solved by
the inflation stage. But the problem with ``washing out'' of
``unneeded objects'', like monopoles, remains.
 The initial and boundary conditions in such a model are set at $t
= - \infty$, and so this can naturally lead to almost arbitrary
density perturbations, which can penetrate trough the $(-t_0) \div
(t_0)$ interval and initiate various structures at the FRW stage
$(t_0) \div \infty$.
 Here on can meet some troubles, because the gravitational
instabilities can grow on the implosion stage $ -\infty \div
(-t_0)$. As a result the perturbations at $t \sim -t_0$ can become
to big, so that $\delta\rho/\rho \sim 1$ on all scales for rather
general initial conditions at  $t = -\infty$. But, in the
high-temperature time interval $(-t_0) \div (t_0)$ the density
irregularities can essentially dump, possibly up to needed level.
 This question needs additional consideration to make any
quantitative estimates.

\subsection{Cosmology with two gravities}

Here we have much more possibilities. The system becomes interesting
when we put a limitation on the geometrical planck scale
$M_P$ such that~:
$$
M_P ~\gg~ \Lambda  ~\sim~ m_p ~\sim~ 10^{19}~ GeV
$$
then the $<\tau_{\mu\nu}>$~-~ condensate scale $\Lambda$ is located
in ``field theory region'', far from $\Lambda_s$~.
Therefore the geometrical
fluctuation of metric near $\Lambda$ are small
\footnote{
This scale ordering looks even natural in such a type models. The
logarithmic grow of gauge constants in $(B,\chi,\varphi)$ field
sector is slow and on can need many orders in scale before on
reaches the distances $\Lambda^{-1}$ when interactions become
strong.}.

It is also favorable to chose the $L(B,\chi,\varphi)$ so that
the physics is supersymmetric from planck(string) scale $\Lambda_s$
up to $\Lambda$. ~This by itself is a natural long range situation
in string theory.
Then there will be no contribution to the geometrical cosmological
constant from distances $(M_P^{-1} \div \Lambda^{-1})$.

From the other hand, all contributions to the mean $<T_{\mu\nu}>$,
coming from scale $\Lambda^{-1}$ and all larger distances,
don't interacts with the goldstone gravity. But in the same time
they look like a cosmological constant for the geometrical gravity.
So we can ``adjust'' the  $M_P$-scale  such
\footnote{
This corresponds to the choice of $L(B,\chi,\varphi)$ parameters
on the $M_P$ scale.
And because this choice of $L$ is not ``free'' and is  in fact
defined by the dynamics of strings probably all such an adjusting
can only be considered as corresponding to the `anthropological'
selection of successful string vacuum.
}
that the mean $\langle T_{\mu\nu}\rangle $ gives no more than the
now ``observable'' cosmological constant
$\rho_{exp} \sim \big( 10^{-3} eV \big)^4$.
This corresponds to condition on $\Lambda$ and~$M_P$~:
$$
\rho_{exp} ~\sim~ \Lambda^4 \Big(\frac{\Lambda}{M_P} \Big)^2~~,
$$
which gives
\bel{planck2}
M_P  ~=~m_p ~\xi  ~\sim~ 10^{81}GeV~,~~~~~~
\xi = M_P/m_p ~\sim \Big(\frac{m_p}{\rho_{exp}^{1/4}}\Big)^2
~\sim~(10^{31})^2 GeV
\ee
for the geometrical Planck (string) scale $M_P$.

Note that this simplest construction (although it can ``explain''
the value of cosmological constant) is not fully perfect,
because in this case the supersymmetry is supposed to be broken
already at $\sim \Lambda$ i.e. at $10^{19} GeV$.
We will have  more preferable picture if supersymmetry
is broken at much lower energy scale $\mu \ll m_p$.
In this case to transform this $\mu$ to the cosmological constant
$\rho_{exp}$ we need string planck mass $M_s$ smaller
then in (\ref{planck2})~:
\bel{scale3}
M_P ~=~ m_p~ \xi ~\sim~ 10^{63}GeV~,~~~~~~~
\xi ~\sim~  \Big(\frac{\mu}{\rho_{exp}^{1/4}}\Big)^2
    ~\sim~ 10^{44}~,
\ee
where the numbers are shown for $\mu \sim 10^{10} GeV$.

\vspace{2mm}

In brane-like models on can try to make the geometrical
planck mass $M_p$ even less.
The actual value depends from a connection between $\Lambda$,
the brane scale $\kappa$ and the ``size'' of large dimensions
$\varepsilon^{-1}$.
We have two possibilities.

If the goldstone gravity is localized on the 3-brane, and the
geometrical gravity can propagate also in  bulk,
then $\kappa > \Lambda$.
The mass $M_P$ can be reduced by the factor
$\eta = \big( \kappa/\varepsilon\big)^{d_{\bot}}$, where
$d_{\bot}$ is the number of ``large'' dimensions~;~ so here
\bel{plmass1}
M_P ~\sim~ m_p~ \xi /\eta~  ~\sim~ 10^{63}/\eta ~GeV~~~~~
\mbox{for}~\mu\sim 10^{10} ~GeV
\ee
where $\eta$ can vary in wide interval but with conditions that
$M_P > \Lambda$ ~-~ this corresponds to $\eta < 10^{44}$.
For $\kappa \sim \Lambda \sim m_p$ this gives the estimates of
size of large dimensions
$$
  \varepsilon^{-1} ~\sim~ \kappa^{-1}~\eta^{1/d_{\bot}} ~<~
    \kappa^{-1}~\xi^{1/d_{\bot}} ~\sim~
10^{-28} \xi^{1/d_{\bot}} ~eV^{-1} ~\Rightarrow~
10^{-28+43/d_{\bot}}
$$
where the last estimate is for $\mu \sim 10^{10} GeV$.

\subsection{Inflation in bigravity models}

Now briefly consider the specifics of inflation in such bigravity
models. Here as in ``usual'' case an inflation can start from the
large and slowly varying coherent fluctuations of some appropriate
``inflaton'' scalar $\phi_{in}$ field. The goldstone gravity do
not interact with such coherent fields - their energy-momentum
tensor is almost $\sim \eta_{\mu\nu}$ inside such fluctuation -
and so does not leads to inflative expansion~
\footnote{But it can interact with the surface of such a bubble,
where the energy density changes, and can lead to a fast collapse
of this fluctuation.}.

But inflation can be caused by the coupling of the geometrical
gravity field to such a big fluctuation of this inflaton field
$\phi_{in}$ arising on the energy scale $\Lambda_{in}$. If the
energy density inside such a fluctuation is $\Lambda_{in}^4$, then
the corresponding de-Sitter expansion stage is characterized by
the scale factor $a(t) \sim\exp (H \delta t)$, where the mean
$$
H~\sim~\Lambda_{in}^2/M_P~=~m_p~(\Lambda_{in}/m_p)^2/\xi ~~,
$$
and the relaxation time $\delta t$ of $H$ depends from a model for
$\phi_{in}$.
Thus to have big value of $H  \sim 10^{12} GeV$, as supposed in
many approaches, we need $\Lambda_{in} ~\sim~ m_p~10^{8 \div 9}$.
Such $\Lambda_{in}$ is still $\ll M_P$ given by (\ref{scale3}) and
where the locale physics corresponds to renormalized field theory.
But from the other side, such a fluctuations are very big for the
$\Lambda$ scale physics, and therefore very improbable. Possibly
some more complicated mechanisms (hybrid , etc) can cure this
\footnote{
Note that to have such an inflation in the case of brane world
with large dimensions we have limitation on the parameter
$\eta < \xi (m_p/\Lambda_{in})$.}.

What concerns the spectrum of initial density perturbations
generated during such inflation stage, with  $\Lambda_{in}
>\Lambda $, they are prepared in pre-FRW stage. And so they will
look in the goldstone gravity mediated FRW epoch as coming
directly from the big-bang~
\footnote{ It is possible that the fluctuations corresponding to
the quadrupole harmonics in CMB can be additionally affected, du
to the interference with the vacuum tensor components
$n_{\mu\nu}$. We thank to F. Viviani for this remark.}.

The general picture of cosmological history in such a bigravity
models can look as the following.
 We have the slow permanent de-Sitter expansion caused by the
geometrical component of gravity - their rate is defined by the
value of $\Lambda^6/\Lambda_{in}^2$ - and corresponds to the
$H=\dot{a}/a$ of order of ``modern time'' experimental  value.

This de-Sitter background is very rarely populated by the bubbles
(Universes like ours) of normal $L(B,\chi,\varphi)$ mater. These
bubbles are created  by the extremely rare inflation-fluctuation
on scales $\Lambda_{in} \gg \Lambda$ as discussed above~
\footnote{Because the probability of inflation initializing
fluctuation can be very small, there will be no more than one such
island inside the primary de-Sitter horizon}.

Inflation with high $H$ is the first fast stage of grow of these
matter-islands and it is driven by the geometrical gravity.
 Then follows the thermalization and the creation of matter with
density and temperature possibly exiting $\Lambda$.
 After that this bubble enters the ``weak'' FRW stage of almost
free expansion until their temperature $T$ is greater than
$\Lambda$.

Further, when temperature becomes less then $\Lambda$, goldstone
gravity with strength $G =m_p^2 \sim \Lambda^2$ switches on, and
the expansion enters the ``normal'' FRW phase, with the rate
defined entirely by the goldstone gravity.
 This will take place up to times when the mean falling matter
density becomes of the order of primary effective cosmological
constant.

After that the de-Sitter expansion of bubbles, driven by the
geometrical gravity, becomes more essential (fast)~
\footnote{In such a model this epoch corresponds to the ``today''
of our Universe. } and they gradually dissolve and disappear.

In fact such type of cosmological scenario can be ``embedded'' in
almost every bigravity theory with strongly separated values of
gravitational constants and if for the more long-range gravity
component the cosmological constant is cancelled by some effective
mechanism. One can make future speculations and include this
scenario in various multiverse models in which only some branches
are equipped with the goldstone-gravitational physics.

\section{\bf Concluding remarks}

The composite goldstone gravity is an interesting possibility which
 could open new ways, especially in cosmological aspects.
It gives rise to a number of interesting questions with can be
discussed for finding a different insight to the existing problems.
These include the structure of horizons and various solitonic objects
like black holes.

Trying to explore the goldstone gravity we encounter serious
problems, and it is yet unclear if they can be solved in a
appropriate way.
 These questions are purely technical, in spite of their
technical complexity, and we hope that they can be
answered somehow.

Evidently, the main question is if it is possible to organize the
dynamics of tensor condensates in such a way to make the breaking
of LI ``invisible'' with a sufficient precision.~
 Moreover, one also needs a better understanding when and how
the tensor condensates can form in the strong coupled gauge
theories. Probably, this can be established by the lattice
stimulation methods.

\vspace{5mm}
{\bf Acknowledgements} \\

\nin We would like to thank J.~Chkareuli, D.~Comelli, V.~Ksenzov,
 S.~Nussinov and A.~Vainstein for useful discussions and comments.
The work was partially  supported by the MIUR biennal grant
PRIN 2006 for the Research Projects of National Interest
on "Astroparticle Physics" and by the FP6 Network "UniverseNet"
MRTN-CT-2006-035863.
O.V.K. thanks the Gran Sasso National Laboratories for hospitality
during the completion of this work  and
RFBR for  the financial support  through
the grants 06-02-72041.


\end{document}